\documentclass[journal=jctcce,manuscript=article]{achemso}

\usepackage[version=3]{mhchem} 
\usepackage{amsfonts}
\usepackage{amssymb,amsmath}
\usepackage[linesnumbered,boxed]{algorithm2e}
\usepackage{enumerate}
\usepackage{bbm}
\usepackage{color}
\usepackage{braket}
\usepackage{subcaption}

\author{Sheng Guo}
\author{Zhendong Li}
\author{Garnet Kin-Lic Chan}\email{gkc1000@gmail.com}
\affiliation{Division of Chemistry and Chemical Engineering,
California Institute of Technology, Pasadena, CA 91125, USA}

\title[\texttt{achemso} demonstration]
{A Perturbative Density Matrix Renormalization Group Algorithm
for Large Active Spaces}

\setkeys{acs}{articletitle=true}

\begin{document}

\begin{abstract}
We describe a low cost alternative to the standard variational
DMRG (density matrix renormalization group) algorithm that is analogous to the combination of selected configuration interaction
plus perturbation theory (SCI+PT).
We denote the resulting method p-DMRG (perturbative DMRG) to distinguish it from the standard variational DMRG.
p-DMRG is expected to be useful for systems with very large active spaces, for which  variational DMRG becomes too expensive. Similar to SCI+PT,
in p-DMRG a zeroth-order wavefunction is first obtained by a standard DMRG calculation, but with
a small bond dimension. Then, the residual correlation is recovered by a second-order
perturbative treatment. We discuss the choice of partitioning for the perturbation theory, which is crucial
for its accuracy and robustness. To circumvent the problem
of a large bond dimension in the first-order wavefunction,
we use a sum of matrix product
states (MPS) to expand the first-order wavefunction, yielding
substantial savings in computational cost and memory.
We also propose extrapolation schemes to reduce the errors in the zeroth- and first-order wavefunctions.
Numerical results for \ce{Cr2} with a (28e,76o) active space and
1,3-butadiene with a (22e,82o) active space reveal that p-DMRG
provides ground state energies of a similar quality to
variational DMRG with very large bond dimensions, but
at a significantly lower computational cost. This suggests that p-DMRG will be an efficient tool
for benchmark studies in the future.
\end{abstract}

\section{Introduction}

Achieving chemical accuracy (ca. 1m$E_h$) in systems with
a mix of multireference and dynamic correlations remains a challenging problem in molecular
quantum chemistry. While complete active spaces (CAS) with tens of partially filled
orbitals can be reliably treated by techniques such as the density
matrix renormalization group (DMRG), \cite{white_density_1992,white_density-matrix_1993, white_ab_1999,mitrushenkov2001quantum,
  chan_highly_2002,legeza2003controlling,sharma_spin-adapted_2012,roberto,keller_efficient_2015,yanai_density_2015,chan_matrix_2016}
reaching chemical accuracy in the subsequent description of the dynamic correlation is difficult.
The most common technique to treat dynamical correlation in the multireference setting is second-order perturbation theory (PT)~\cite{andersson_second-order_1990, roos_multiconfigurational_1996,angeli_introduction_2001,angeli_n-electron_2001, angeli_n-electron_2002, kurashige_second-order_2011, sharma_communication:_2014, guo_n-electron_2016, Alex_t_nevpt, Reigher_dmrg_nevptpt, Alex_t_dmrg_pt, Naoki_caspt2}.
However, one often finds that a second-order perturbative treatment is not powerful enough
to accurately describe correlations involving some of the moderately correlated non-valence orbitals in a complex system. For example,
in $3d$ transition metal systems, binding energies and exchange couplings can be substantially in error
if the virtual $4d$, semi-core $3s3p$, or valence ligand orbitals, are treated only at the second-order perturbative level.
The standard remedy is to include these additional moderately correlated orbitals in
the multireference active space treatment. However, for complex systems this can create
 enormous active spaces that are inaccessible or otherwise impractical even for current DMRG methods.

Recently, selected configuration interaction (SCI) methods\cite{huron1973iterative,buenker1974individualized,harrison1991approximating}
have experienced a significant revival~\cite{schriber2016communication,
  tubman2016deterministic,liu2016ici,holmes_heat-bath_2016,sharma2017semistochastic,garniron2017hybrid}.
The general idea of selected configuration interaction is quite old, dating back to the 
CIPSI method~\cite{huron1973iterative}, and before
that, to the hand-selected configuration interaction calculations carried 
out in the earliest days of quantum chemistry~\cite{foster1960canonical,bender1969studies}.
Although modern day SCI methods differ in how they select determinants, they all share a similar basic strategy.
In particular, a small number of determinants are first selected for a variational treatment - in modern calculations, typically $10^6$-$10^7$ determinants -
and the residual correlation is treated by second-order PT, most commonly using the Epstein-Nesbet (EN)
partitioning. Some important recent improvements include the use of stochastic methods to evaluate the
second-order energies ($E_2$) in order to handle large basis sets\cite{sharma2017semistochastic,garniron2017hybrid}, as
well as the development of more systematic extrapolations with respect to the thresholds in the method.
One finds that SCI methods achieve chemical accuracy in the total energy for a variety of small molecule problems using a remarkably small number of variational determinants.
However, it is important to observe that the variational CI energy alone is itself usually quite poor. For example, in a heat-bath CI calculation
on the chromium dimer (48e, 42o) active space~\cite{holmes_heat-bath_2016} popularized in DMRG benchmarks~\cite{roberto}, the variational CI energy was more than 60 m$E_h$ above the
the DMRG benchmark result. Instead, it is the second order PT correction, combined with extrapolation,
that yields the final high accuracy result. In the above case, the total energy error using perturbation theory plus extrapolation
is reduced to less than 1 m$E_h$, a reduction by a factor of almost one hundred.

The remarkable accuracy of the second-order perturbation correction in selected CI stands in
stark contrast to the accuracy of second-order perturbation corrections when used with complete active spaces. The physical reason for
the difference is that even if the reference wavefunction is determined exactly (within the complete active space)
it is {\it unbalanced} due to the lack of dynamical correlation.
In contrast, although the variational selected CI computes only a quite approximate
reference wavefunction, it is determined in a full, or at least large, space of orbitals, leading to a
more balanced reference state. This suggests that the key to an accurate second-order correlation contribution
involves balancing the different orbital correlations, rather than describing only the strongest correlations exactly, as in a valence CAS.
This observation is independent of choosing selected CI for the reference wavefunction,
and it is the motivation for this work.

In the current paper, we will explore how we can use quite approximate, but balanced, variational DMRG reference wavefunctions
computed in large active spaces, and correct them efficiently and to high accuracy, with second order PT within the same orbital space.
We name this technique ``perturbatively corrected DMRG'' or p-DMRG.
In p-DMRG, we represent both the zeroth order variational reference wavefunction $|\Psi^{(0)}\rangle$ as well as the first order perturbative correction $|\Psi^{(1)}\rangle$
in terms of matrix product states (MPS). Note that there are advantages to using a MPS representation, rather
than a determinantal expansion, of the variational reference wavefunction. The MPS representation
allows us to construct compact strongly correlated wavefunctions even where there is little to no determinantal sparsity,
for example in systems with many coupled spins, where there is little sparsity in the coupled low-spin configurations of the system.
A second reason is that volume extensivity of the energy is achieved by a matrix product state
with a cost $\propto e^{V^{2/3}}$ rather than $\propto e^V$ in configuration interaction. Asymptotically, this makes the variational MPS
representation exponentially more compact than a variational determinant expansion, and in practice, allows for a larger number of spatially separated
orbitals to be treated~\cite{Hachmann_Hchain}.

Relative to a standard variational DMRG calculation, the cost savings in p-DMRG arise from two sources.
First, as described above, the zeroth order wavefunction can be computed using a bond dimension $M_0$ much smaller than is needed
to fully converge the variational DMRG calculation. Second, although the bond dimension $M_1$ for the first order wavefunction
still needs to be quite large, the first order wavefunction it is determined by minimizing the Hylleraas functional\cite{Hylleraas_functional,sharma_communication:_2014},
\begin{equation}
\mathcal{L}[\ket{\Psi_1}] = \bra{\Psi_1}(\hat{H}_0 -E_0)\ket{\Psi_1} +2\bra{\Psi_1}\hat{V}\ket{\Psi_0},\quad
\hat{V}=\hat{H}-\hat{H}_0.\label{eq:Hy}
\end{equation}
which is less expensive than minimizing the variational DMRG energy, because the zeroth order Hamiltonian $\hat{H}_0$
can be chosen to be simpler than the full Hamiltonian $\hat{H}$. For example, if $\hat{H}_0$ is
the Fock operator or the Epstein-Nesbet Hamiltonian, then the computational cost to evaluate the Hylleraas functional
is a factor of $K$ (where $K$ is the number of orbitals) less than that to evaluate the variational DMRG energy.
In addition, since in second-order PT, only the matrix element $\bra{\Psi_1}\hat{V}\ket{\Psi_0}$ needs to be computed (instead of $\bra{\Psi}\hat{H}\ket{\Psi}$ in standard DMRG)
we can save a further factor of $M_1/M_0$ in cost, where we assume $M_1$ is similar to the bond dimension used in a converged variational DMRG
calculation, and $M_1 \gg M_0$.
The p-DMRG method can still be made exact by gradually increasing $M_0$,
which thus plays a role analogous to the variational selection threshold in SCI methods.
This opens up the possibility to perform extrapolations, similarly to as done in SCI and
in variational DMRG.

It is important to note that we expect p-DMRG
to be useful for a different class of problems than
 standard DMRG based multi-reference perturbation theory such as DMRG-CASPT2~\cite{kurashige_second-order_2011, Naoki_caspt2} or DMRG-NEVPT2~\cite{sharma_communication:_2014, guo_n-electron_2016, Reigher_dmrg_nevptpt, Alex_t_dmrg_pt}.
In particular, we believe the method should be used to target high accuracy
calculations (to say 1m$E_h$ in the total energy) either in a  large active space, including the intermediately correlated orbitals, or to
obtain benchmark total energies in small problems, at a cost that is significantly less than that of variational DMRG. This
is very different from providing a qualitative treatment of dynamical correlation in very large basis sets, which is
the focus of standard DMRG based multi-reference PT.
Note that p-DMRG differs also from the similarly named DMRG inner space perturbation theory (DMRG-isPT)\cite{ren_inner_2016},
where the PT is only used to
reduce the cost of the Davidson diagonalization in the DMRG sweeps.

The remainder of the paper is organized as follows. In Sec.
\ref{theory:pdmrg}, we first briefly summarize DMRG in the MPS language
and then introduce the p-DMRG algorithm.
Two particular pieces needed to establish p-DMRG as an accurate
and efficient alternative to variational DMRG are then discussed in the following sections. Specifically, Sec. \ref{theory:H0} discusses the choice of $\hat{H}_0$, which is crucial for
obtaining high accuracy, while Sec. \ref{theory:psi1} introduces a way to tackle the
large bond dimension $M_1$ needed to represent the first order wavefunction by using a sum
of MPS representations.
After describing standard benchmark calculations for \ce{C2} and \ce{Cr2}
in small active spaces, we carry out two larger benchmark studies  using p-DMRG in Sec. \ref{results}:
one for \ce{Cr2} in an active space with 28 electrons in 76 orbitals generated by a cc-pVDZ-DK basis, denoted by
the notation (28e,76o), and the other for butadiene in an active space with (22e,82o) generated
by a cc-pVDZ basis.
Both sets of calculations demonstrate that in practical problems p-DMRG is substantially more efficient than variational DMRG,
obtaining the same benchmark accuracy with greatly reduced cost. Conclusions and outlines for future directions are presented in Sec. \ref{conclusion}.

\section{Theory}\label{theory}
\subsection{Perturbative density matrix renormalization group (p-DMRG)}\label{theory:pdmrg}

Here we first recapitulate the DMRG algorithm in the MPS language.
Interested readers are referred to recent reviews, e.g., Refs. \cite{schollwock_density-matrix_2011,keller2016spin,chan_matrix_2016}
for details.

A generic FCI wavefunction can be written in Fock space as
\begin{equation}
  \ket{\Psi} = \sum_{n_1\cdots n_K}\Psi^{n_1 n_2 \cdots n_K} \ket{n_1 n_2 \cdots n_K},
\end{equation}
where $\ket{n_1 n_2\cdots n_K}$ is the occupation basis in the Fock space of $K$ spatial orbitals,
and $n_k\in\{0,1,2,3\}$ for the local configuration basis $\{|0\rangle,|k_\beta\rangle,|k_\alpha\rangle,|k_\alpha k_\beta\rangle\}$, respectively.
It can be decomposed into a sequential product of matrices associated with different orbitals
via successive singular value decompositions (SVDs),
\begin{equation}
  \ket{\Psi} = \sum_{n_1\cdots n_K}A^{n_1}[1] A^{n_2}[2]\cdots A^{n_K}[K]\ket{n_1 n_2\cdots n_K},\label{eq:MPS}
\end{equation}
where $A^{n_k}[k]$ are matrices and the symbol $A[k]$ will be used to represent
the site tensor as a collection of matrices $A^{n_k}[k]$ for different $n_k$.
The dimensions of $A^{n_k}[k]$ are usually referred to as the bond dimensions,
and these take a maximal value of $O(4^{K/2})$ in the middle of the orbital chain\cite{schollwock_density-matrix_2011}.
The MPS form \eqref{eq:MPS} can be used as a variational
ansatz by restricting the maximal bond dimension
to a given $M$, which is then the single parameter that controls the accuracy of the approximation.
Clearly, as $M$ approaches $O(4^{K/2})$, the ansatz  becomes exact.
However, the importance of the MPS ansatz is that for Hamiltonians with local interactions
in one dimension, the entanglement encoded in an MPS with an $M$ with only a
very weak dependence on $K$, is sufficient to accurately represent  ground and low-energy eigenstates. For real molecules which have a more complicated
entanglement structure, the required $M$ is generally much larger than that used in one-dimensional models~\cite{chan_highly_2002}.

The DMRG algorithm provides an efficient way to variationally optimize an MPS that optimizes the tensors site-by-site. For simplicity, we consider here only the single site sweep algorithm.
When we optimize the site tensor $A[k]$ at site $k$, the MPS can be recast into a \emph{mixed-canonical} form as
\begin{equation}
  \ket{\Psi} = \sum_{n_1\cdots n_K}L^{n_1}[1]\cdots L^{n_{k-1}}[k-1]C^{n_k}[k]R^{n_{k+1}}[k+1]\dots R^{n_K}[K] \ket{n_1n_2\cdots n_K},
\end{equation}
where the set of $L[k]$ are in left canonical form ($\sum_{n_k}L^{n_k\dagger} L^{n_k} = I$)
and the set of $R[k]$ are in right canonical form ($\sum_{n_k}R^{n_k} R^{n_k\dagger} = I$).
This choice of the left and right canonical gauges makes the renormalized configuration
basis $\{|l_{k-1}n_k r_{k}\rangle\}$ orthonormal, where
\begin{eqnarray}
|l_{k-1}\rangle &=& \sum_{n_k} (L^{n_1}[1]\cdots L^{n_{k-1}}[k-1])_{l_{k-1}}\ket{n_1\cdots n_{k-1}},\\
|r_{k}\rangle &=& \sum_{n_k} (R^{n_{k+1}}[k+1]\dots R^{n_K}[K])_{r_{k}}\ket{n_{k+1}\cdots n_{K}},
\end{eqnarray}
that is, $\langle l_{k-1}'|l_{k-1}\rangle=\delta_{l_{k-1}' l_{k-1}}$ and
$\langle r_{k}'|r_{k}\rangle=\delta_{r_k'r_k}$. The central part $C^{n_k}[k]$
is the wavefunction to be optimized at site $k$, and it can be obtained by solving
a standard configuration interaction problem in the renormalized configuration basis
$\{|l_{k-1}n_k r_{k}\rangle\}$,
\begin{equation}
\sum_{lnr} H_{l'n'r',lnr}[k] C_{lnr}[k] = E C_{lnr}[k],
\end{equation}
where $H_{l'n'r',lnr}=\langle l_{k-1}'n_k'r_k'|\hat{H}|l_{k-1}n_kr_k\rangle$ is the matrix
representation of the Hamiltonian $\hat{H}$ and
$C_{lnr}[k]$ is the vectorized version of the tensor $C^{n_k}_{l_{k-1}r_{k}}[k]$.
The multiplication between $H_{l'n'r',lnr}$ and $C_{lnr}$ dominates
the cost of a DMRG calculation, and scales as $O(K^3M^3)$ in total per sweep\cite{white_ab_1999,chan_highly_2002}.
This scaling can be understood by noting that $H_{l'n'r',lnr}$ can always be written
as a sum of $O(K^2)$ direct product terms $H_{l'n'r',lnr}=\sum_{\beta}O^{\beta}_{l'n',ln}O^\beta_{r'r}$ for a generic second
quantized Hamiltonian with $O(K^4)$ terms, via the complementary operator technique\cite{xiang1996density,white_ab_1999,chan_highly_2002},
such that the matrix vector product can be formed by $O(K^2)$ independent matrix multiplications
\begin{eqnarray}
\sigma_{l'n'r'}=\sum_{lnr}H_{l'n'r',lnr}C_{lnr}=
\sum_{\beta}\left(\sum_{r}
\left(\sum_{ln}
O^{\beta}_{l'n',ln}C_{lnr}\right)O^\beta_{r'r}\right).
\end{eqnarray}
The  cost for each multiplication scales as $O(M^3)$, thus
the cost for forming $\sigma_{l'n'r'}$ scales as $O(K^2M^3)$ at a given site $k$.
In combination with the cost for building the necessary operators $O^{\beta}_{l'n',ln}$ and
$O_{r'r}^\beta$ for representing $H_{l'n'r',lnr}$, the computational cost for
the standard DMRG algorithm using the quantum chemistry Hamiltonian scales as $O(K^3M^3+K^4M^2)$,\cite{white_ab_1999,chan_highly_2002} which, unlike FCI, is a polynomial in $K$, if $M$ can be kept constant as a function of $K$,
as is the case in certain situations, such as in pseudo-one-dimensional molecules.

However, to describe dynamical correlation in a small molecule over length scales too short for  locality of correlations to emerge,
$M$ needs to scale as $O(K)$ to capture the local double excitations~\cite{roberto}. This renders the total scaling effectively $O(K^6)$.
This limits the number of orbitals that can be treated accurately with reasonable computational resources and time. For instance, as shown in Ref. \cite{roberto}, a state-of-the-art DMRG calculation on
butadiene with an active space (22e,82o) took one day on 42 cores for a single sweep with $M=3000$. In this scenario, the correlation treatment
offered by the MPS, where every orbital is treated on an equal footing, is too flexible.
Thus, a less general, but more efficient formulation, is clearly desired.

In the p-DMRG method, we assume that an MPS with small $M_0$ has been optimized
by the above standard DMRG algorithm, and it is used as the zeroth-order
wavefunction $|\Psi^{(0)}\rangle$. Then, the first-order wavefunction
$|\Psi^{(1)}\rangle$ can be obtained by minimizing the Hylleraas functional \eqref{eq:Hy},
which in the exact case is equivalent to solving the first-order equations,
\begin{equation}
(\hat{H}_0 -E_0)\ket{\Psi^{(1)}} = -Q\hat{H}\ket{\Psi^{(0)}},
\quad Q=1-\ket{\Psi^{(0)}}\bra{\Psi^{(0)}}.\label{eq:linear}
\end{equation}
Note that although the bond dimension of $|\Psi^{(0)}\rangle$ is chosen small, the bond dimension $M_1$ of $|\Psi^{(1)}\rangle$ arising
from \eqref{eq:linear} can
 be substantially larger, for example, as large as the bond dimension used in a converged variational DMRG calculation.
 In the following sections, we will discuss different definitions of the zeroth-order Hamiltonian $\hat{H}_0$, and how to solve
 the first-order equation efficiently for the large bond dimensions arising in $|\Psi^{(1)}\rangle$.

\subsection{Choices of zeroth-order Hamiltonian $\hat{H}_0$}\label{theory:H0}
There are several criteria that a good partitioning of $\hat{H}$ must satisfy.
First, in order to reduce the computational cost, $\hat{H}_0$ should be as simple as possible.
The Fock operator or the diagonal part of $\hat{H}$ in the determinant space used in the EN partition
both satisfy this criteria, while the simplest projective definition $\hat{H}_0= P\hat{H}P+Q\hat{H}Q$ does not.
Second, the partition should be free of intruder state problems. The Fock operator generally does not satisfy this
criterion (as we have numerically verified) and hence will not be discussed further.
Instead, we will exclusively focus on  designing $\hat{H}_0$ based
on the idea of the EN partition, as also used in SCI+PT schemes\cite{huron1973iterative,buenker1974individualized,harrison1991approximating,
schriber2016communication,tubman2016deterministic,liu2016ici,holmes_heat-bath_2016,sharma2017semistochastic,garniron2017hybrid}.
Third, the partition should give good energies at 2nd order, which requires a balanced treatment of $|\Psi^{(0)}\rangle$
and $|\Psi^{(1)}\rangle$.
Fourth, to be used in a spin-adapted DMRG algorithm\cite{sharma_spin-adapted_2012},
we require a spin-free $\hat{H}_0$. This differs from the partitioning in
determinant based SCI+PT, where $\hat{H}_0$ does not commute with
the spin squared operator $\hat{S}^2$, and leading to spin contamination in the first-order wavefunction.

To begin, we start with $\hat{H}_0$ defined as
\begin{equation}
  \hat{H}_0 = P E_0 P + Q\hat{H}_d Q,\label{eq:H0}
\end{equation}
where $\hat{H}_d$ contains all single and double excitations which do not change the occupation numbers of spatial orbitals,
\begin{equation}
  \hat{H}_d = \sum_{i} h_{ii} \hat{E}_{ii} + \frac{1}{2}\sum_{i,j} (ii|jj)\hat{e}_{ijji} + \frac{1}{2}\sum_{i \neq j} (ij|ji)\hat{e}_{ijij},\label{eq:Hd}
\end{equation}
with $\hat{E}_{ij}=\sum_{\sigma}a_{i\sigma}^\dagger a_{j\sigma}$
and $\hat{e}_{ijkl} = \sum_{\sigma,\tau} a^\dagger_{i\sigma}a^\dagger_{j\tau}a_{k\tau} a_{l\sigma}=E_{il}E_{jk}-\delta_{jl}E_{ik}$.
$\hat{H}_0$ defined in \eqref{eq:H0} is analogous to the zeroth order Hamiltonian in the EN partition, but
it is spin-free. A consequence of this is that it is block-diagonal rather than diagonal in the determinant basis, since
it contains additional couplings for determinants with the same
spatial occupations due to the $\hat{e}_{ijij}$ operator in the exchange term.
Numerical comparisons within a non-spin-adapted DMRG
implementation~\cite{li2017spmps} demonstrate that $\hat{H}_d$
and the standard EN partition provide results of very similar quality.
For this form of $\hat{H}_0$, when solving
Eq. \eqref{eq:linear} using the DMRG sweep algorithm, the
Hamiltonian and wavefunction multiplication on the left hand side (LHS)  scales as
$O(K^2M_1^3)$ instead of $O(K^3M^3)$ in the standard variational DMRG.
The construction of the right hand side (RHS) will scale as $O(K^3M_1^2M_0)$
assuming $M_1\gg M_0$. The cost to build the renormalized operators
for the LHS is negligible, as it is only  $O(K^2M_1^2)$, while the corresponding cost
for the RHS is  $O(K^4M_1M_0)$ in total. Thus compared to the variational DMRG calculation with a similar $M\approx M_1$,
we expect a substantial reduction in cost.

In Eq. \eqref{eq:H0}, we have not yet defined the zeroth-order energy $E_0$.
There are two natural choices. One is the DMRG energy for $|\Psi^{(0)}\rangle$, viz., $E_{DMRG}^{(0)}=\langle\Psi^{(0)}|\hat{H}|\Psi^{(0)}\rangle$,
which is analogous to the choice made  in SCI+PT.
However, we observe that, unlike in SCI+PT,
the zeroth order variational energy $E_{DMRG}^{(0)}$ is typically much closer to the exact energy than
the zeroth order energies used in SCI+PT. It is
hence much lower than the lowest energy of the perturbers, which is
the lowest eigenvalue of $Q\hat{H}_dQ$, whose eigenstates are relatively uncorrelated.
Thus, although this choice of $E_0$ is in general numerically stable,
and is free of intruder state problems as long as $M_0$ is large enough
to achieve a non-vanishing gap between the zeroth-order state and the perturbers, the correlation energy recovered is usually too small
at the second order level.
The other natural choice $E_{d}^{(0)}=\langle\Psi^{(0)}|\hat{H}_d|\Psi^{(0)}\rangle$
makes the gap smaller and hence lowers $E_2$,
but in this case the correlation energy can be overestimated and there is a greater probability of intruder states,
because there is no guarantee that the lowest eigenvalue of $Q\hat{H}_dQ$ is larger than $E_{d}^{(0)}$.
Therefore, in general, we expect that an interpolation
$E_0(\lambda)=(1-\lambda) E_{DMRG}^{(0)}+\lambda E_{d}^{(0)}$
between these two limits will provide better performance in terms of stability and accuracy.

\begin{figure}
\begin{tabular}{cc}
    \includegraphics[width=0.45\textwidth]{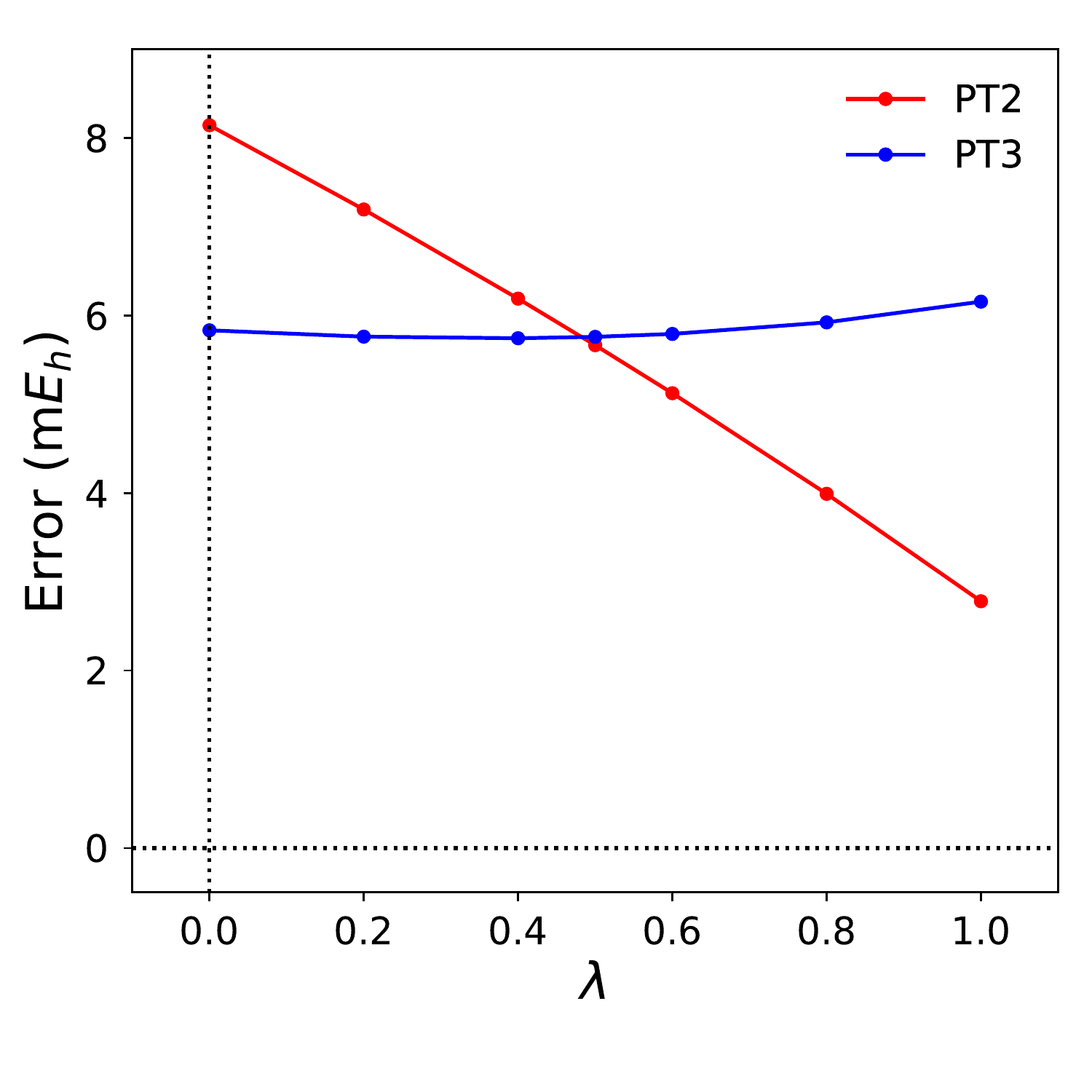} &
    \includegraphics[width=0.45\textwidth]{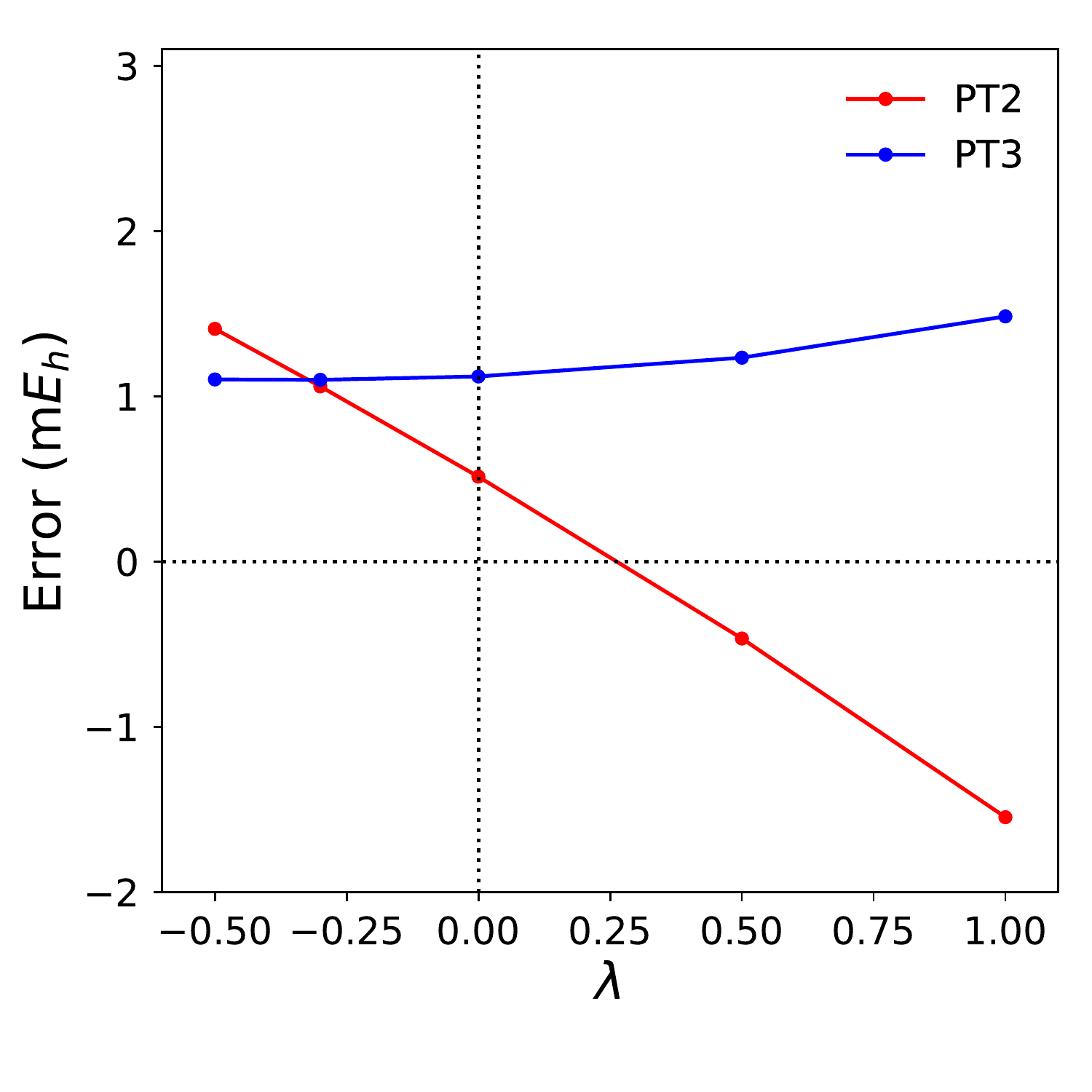}  \\
    (a) \ce{H10} & (b) \ce{H2O}
\end{tabular}
  \caption{Dependence of the absolute errors of second- and third-order perturbation theories
  on $E_0(\lambda)=(1-\lambda)E_{DMRG}^{(0)}+\lambda E_d^{(0)}$: (a) \ce{H10}
  with $R$(H-H)=1.0{\AA} and $M_0$=12 in a STO-3G basis; (b) \ce{H2O} at the equilibrium geometry
  and $M_0$=10 in {DZ} basis.}\label{fig:lambda}
\end{figure}

Unfortunately, there is no a priori way to determine $\lambda$ without calculation.
One way to define it  through a calculation, is through the optimized
partitioning method\cite{surjan2000optimized},
where $\lambda$ is chosen to make
$E_3(\lambda)=0$ or equivalently
$E_2(\lambda)+E_3(\lambda)$ stationary, while
$E_0(\lambda)+E_{1}(\lambda)=E_{DMRG}^{(0)}$ is
independent of $\lambda$.
We have explored the dependence of the absolute errors of
second- and third-order perturbation theories (PT2 and PT3)
on $\lambda$ as shown in Figure \ref{fig:lambda}
for two small systems, viz., a hydrogen chain \ce{H10}
with $R$(H-H)=1.0{\AA} in a STO-3g basis\cite{doi:10.1063/1.1672392}
and \ce{H2O} at the equilibrium geometry\cite{chan_highly_2002} in the Dunning's DZ basis\cite{doi:10.1063/1.1674408}.
It is clear that as $\lambda$ increases and $E_0(\lambda)$ approaches
$E_d^{(0)}$, $E_2(\lambda)$ is lower, for
the reasons discussed above. In contrast,
the PT3 energy varies more slowly.
However, including PT3 does not always improve the results, e.g., for \ce{H2O}, the error of PT2+PT3
is larger than using PT2 alone when $\lambda=0$.
Empirically, we observe that the error obtained
at the midpoint $\lambda=1/2$ is always improved over that obtained with $\lambda=0$.
Hence, in the following, we will use this simple choice
in addition to the two obvious choices $\lambda=0$ and $\lambda=1$.

\subsection{Splitting the first order wavefunction}\label{theory:psi1}

In general, for large numbers of orbitals, the bond dimension  $M_1$ required to achieve a given
accuracy increases with $K$. Thus, the
dominant scaling when solving for the first-order wavefunction
is dominated by the scaling $O(K^2M_1^3)$ encountered when computing the LHS of
Eq. \eqref{eq:linear}. A similar computational obstacle
arises also in SCI+PT, which gives rise to the memory bottleneck
associated with storing all determinants contributing to the first-order wavefunction.
One way to remove this bottleneck is to use a stochastic computation of the perturbation correction, as proposed in
\cite{sharma2017semistochastic,garniron2017hybrid}
for SCI+PT. In the current work, we will use a deterministic approach, where we represent
the first-order wavefunction as a linear combination of MPS\cite{split_mps}, each with a modest bond dimension.

Specifically, noting that Eq. \eqref{eq:linear} is a linear equation,
we use the following ansatz,
\begin{eqnarray}
|\Psi^{(1)}\rangle=\sum_{i=1}^{N}|\Psi^{(1)}_i\rangle,\label{eq:split}
\end{eqnarray}
where each $|\Psi^{(1)}_i\rangle$ is represented by an MPS
with a fixed bond dimension $M_1$, and can be determined recursively from
the relation
\begin{equation}
  (\hat{H}_0-E_0)\ket{\Psi_i^{(1)}}
  = |r_i\rangle,\quad
  |r_i\rangle = -Q\hat{H}\ket{\Psi^{(0)}} -\sum_{j=1}^{i-1}
  (\hat{H}_0-E_0)\ket{\Psi_j^{(1)}}.
\end{equation}
The form of the LHS is the same for each $i$, but
the RHS becomes more costly as $N$ increases.
When computed from the Hylleraas functional,
the largest cost arises from computing the expectation value
$\bra{\Psi_i^{(1)}}(\hat{H}_0-E_0)\ket{\Psi_j^{(1)}}$ ($i>j$)
and this cost scales as
$O(K^2M^3_1N^2)$. Thus, using the split ansatz \eqref{eq:split},
compared with a calculation using a large bond dimension
$M_1'=NM_1$, formally leads to a factor of $N$ reduction in computational cost,
as well as a factor of $N^2$ in memory.
However, the representational power of an MPS with
$M_1'=NM_1$ is larger than that of a linear combination
of $N$ MPS with bond dimension $M_1$ due to the compressibility of
the sum of MPS representation.
For example, in the limiting case of $M_1=1$,
Eq. \eqref{eq:split} simply becomes
a sum of $N$ determinants, while the variational space
described by MPS with bond dimension $N$
is of course much larger.
Thus, in practice, we try to use an $M_1$ as large as possible
given the computational resources, and only then
use Eq. \eqref{eq:split} to continue the calculations to a larger effective $M_1$, which would otherwise
be too costly within a single MPS representation.
The second order energy $E_{2}=\langle\Psi^{(0)}|V|\Psi^{(1)}\rangle=
\sum_{i=1}^{N}E_{2,i}$
becomes a sum of $N$ terms,
where $E_{2,i}$ decays monotonically
as $i$ increases. This monotonic decay can be quite systematic and we will explore the possibility to
extrapolate the series $\{E_{2,i}\}_{i=1}^{N}$ for large calculations in Sec. \ref{sec:Cr2}.

\section{Results}\label{results}
\subsection{Benchmark: \ce{C2} and \ce{Cr2}}

To test the performance of p-DMRG for various choices of $\hat{H}_0$,
we examined two diatomic molecules: \ce{C2} and \ce{Cr2}, for which
variational DMRG results are available in the literature\cite{roberto}.
The same molecules were also studied in recent Heat-Bath CI plus PT calculations\cite{holmes_heat-bath_2016}.
For these two molecules, we used canonical Hartree-Fock orbitals
with $D_{2h}$ symmetry and ordered them using genetic ordering as used in Ref. \cite{roberto}.
The zeroth-order DMRG wavefunctions were computed in a default forward sweep where
$M_0$ was increased gradually, using the \textsc{Block} code\cite{chan_highly_2002,sharma_spin-adapted_2012}.

\begin{figure}
  \includegraphics[width=0.45\textwidth]{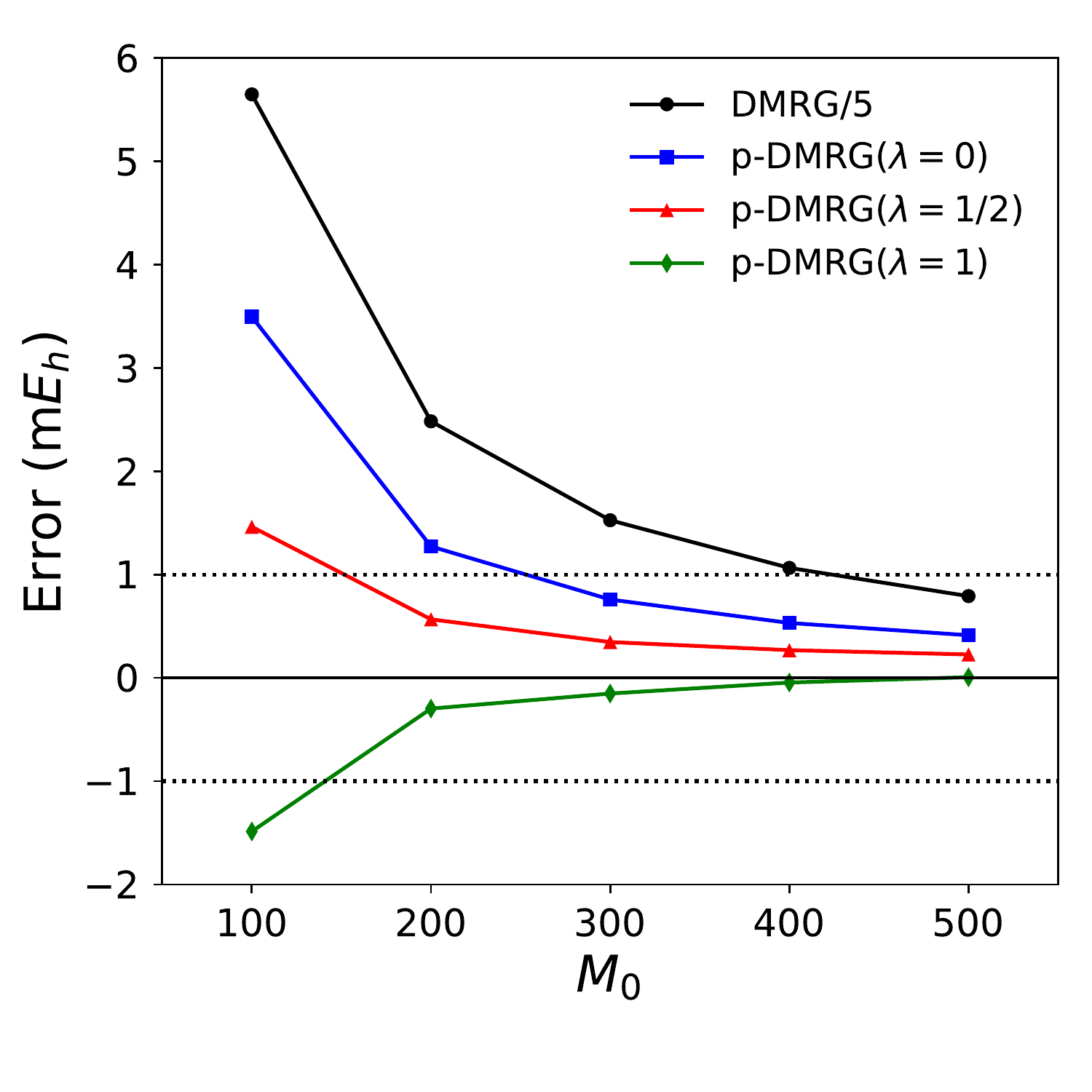}
  \caption{Absolute errors in zeroth-order DMRG energies $E_{DMRG}^{(0)}$ and perturbation corrections $E_{DMRG}^{(0)}+E_2(\lambda)$ with different $\hat{H}_0$. The errors are calculated relative to the converged variational DMRG energy in Ref. \cite{roberto}. The errors of zeroth-order DMRG energies are divided by 5 to put all curves into the same figure.}
  \label{fig:C2}
\end{figure}

Figure \ref{fig:C2} shows the p-DMRG results for \ce{C2} at the equilibrium bond length of 1.24253{\AA} in the cc-pVTZ basis set\cite{doi:10.1063/1.462569}. All electrons were correlated corresponding to an orbital space of (12e,60o). The absolute errors are given relative to the
essentially exact  variational DMRG value\cite{roberto}.
The second order perturbation energies were calculated at an effective $M_1=\infty$ by extrapolating with
discarded weight from $M_1 = 5000, 4000, 3000$ in reverse sweep mode\cite{roberto}.
We first note the significance of the perturbation correction: to compare the variational DMRG
and p-DMRG calculations as a function of $M_0$ on the same plot, we had to divide the variational
error by 5. We also see that in this dynamic correlation dominated system,
the performance of the zeroth order Hamiltonian with  $\lambda=1$ is quite good. Using $E_{DMRG}^{(0)}$ as $E_0$ instead
underestimates the correlation energy. $\lambda=1/2$ also yields reasonable errors
which reach chemical accuracy already for the very small variational DMRG calculation with  $M_0=200$. We see that in the absence
of intruder state problems, p-DMRG with different choices of $\lambda$ all converge
to the same ground state energy as $M_0$ increases, but the accuracy
when $M_0$ is small can be quite different. For this reason, it is important
to choose $\lambda$, such that one obtains good accuracy already with small $M_0$,
to obtain significant computational savings.

\begin{figure}
\begin{tabular}{cc}
    \includegraphics[width=0.45\textwidth]{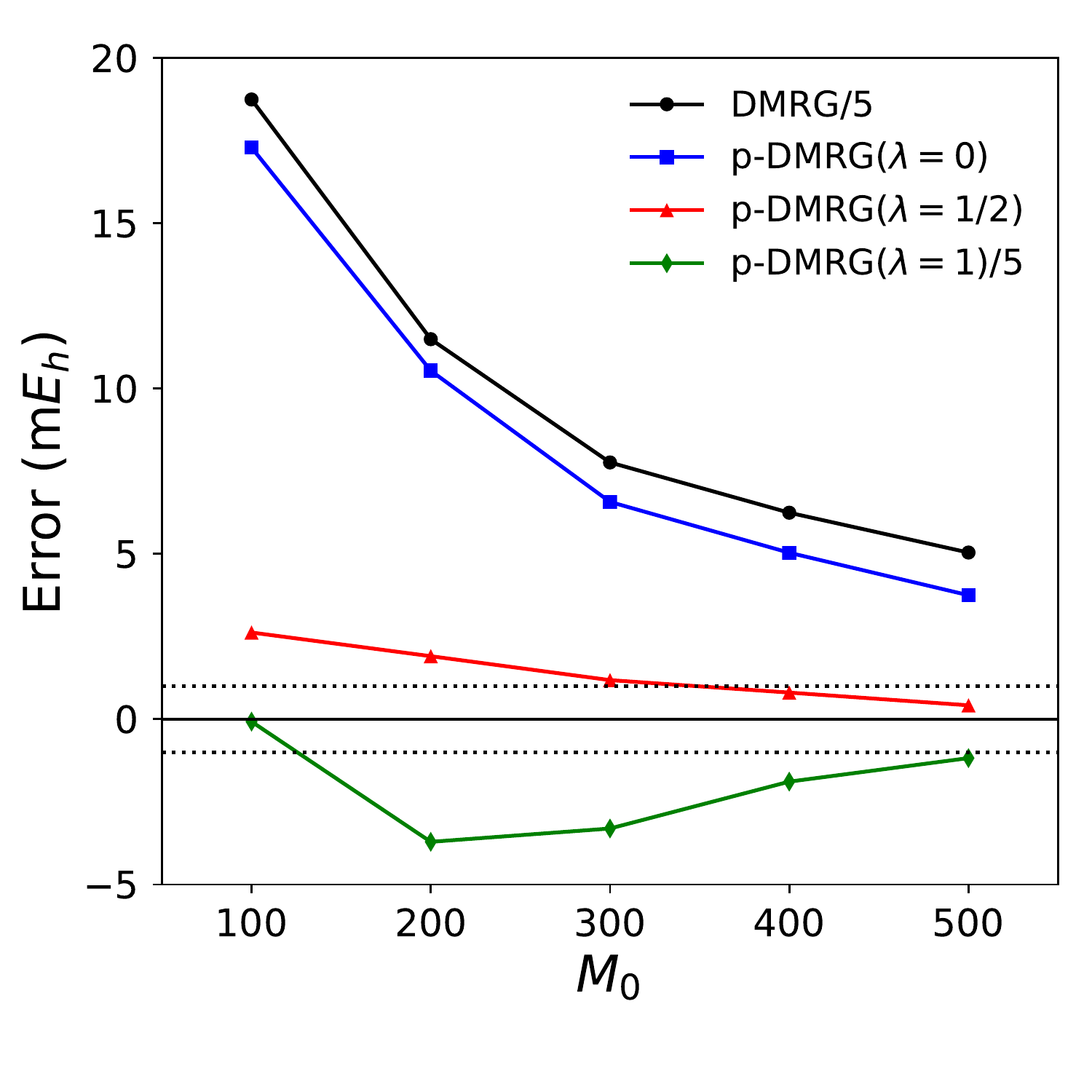} &
    \includegraphics[width=0.45\textwidth]{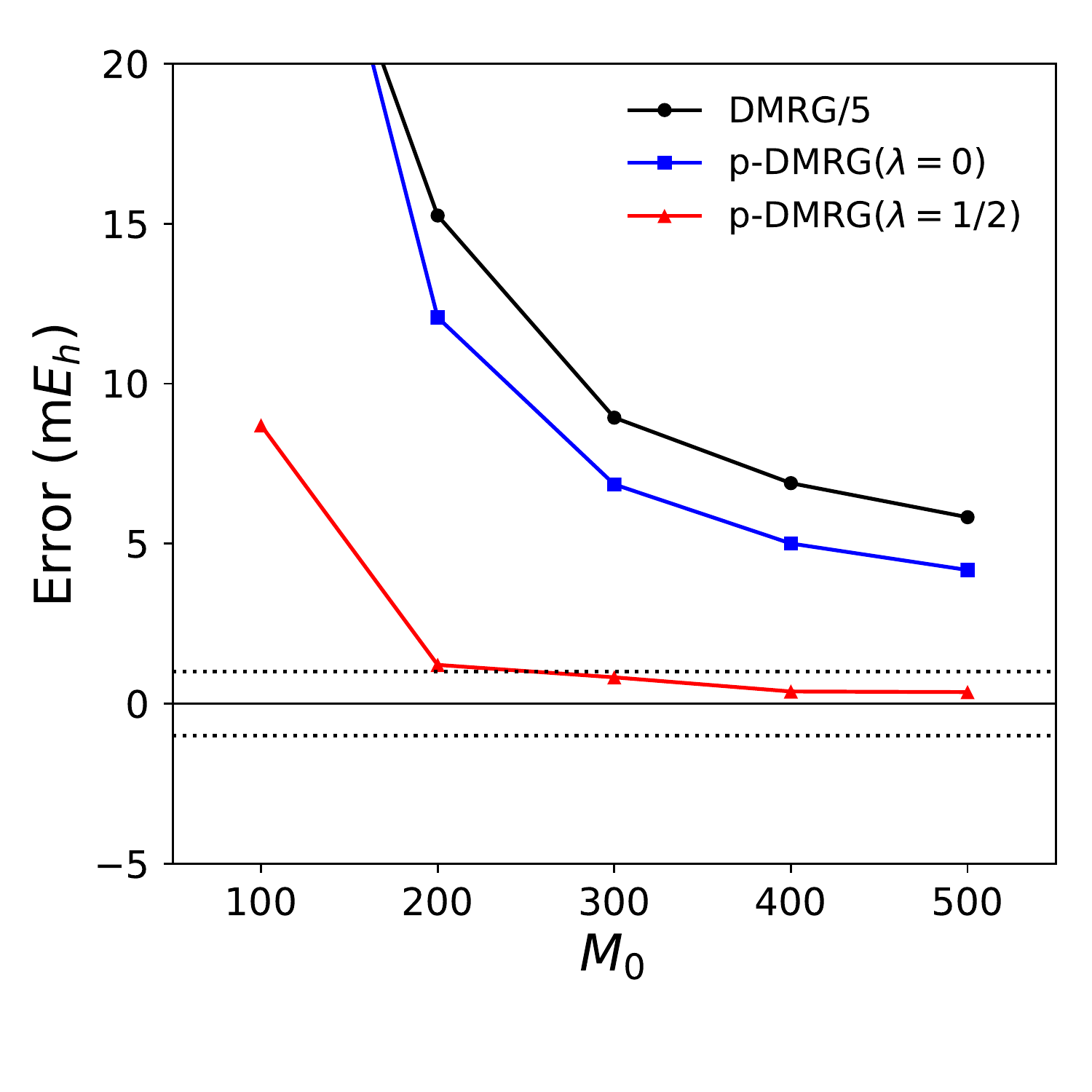}  \\
    (a) $R$=1.68{\AA} & (b) $R$=1.50{\AA}
\end{tabular}
  \caption{Absolute errors in zeroth-order DMRG energies $E_{DMRG}^{(0)}$ and
  perturbation corrections $E_{DMRG}^{(0)}+E_2(\lambda)$ with different $\hat{H}_0$. Both the errors of DMRG energies and p-DMRG$(\lambda=1)$ were divided by 5 to fit all curves on the same figure. For $R$=1.50{\AA}, p-DMRG$(\lambda=1)$ suffers from intruder state problems.}
  \label{fig:Cr2}
\end{figure}

Next, we consider a more challenging example, \ce{Cr2}, at two bond distances,
the equilibrium bond length\cite{bondybey_electronic_1983} $R$=1.68{\AA} and
$R$=1.50{\AA}, which have been previously benchmarked by variational DMRG \cite{roberto}.
We used the Ahlrichs' SV basis set\cite{doi:10.1063/1.463096} and correlated all electrons.
 The resulting orbital space is (48e,42o).
The second order perturbation energies were calculated at $M_1=\infty$ by extrapolation from
$M_1$ = 8000, 7000, 6000 (in reverse sweep mode).
The p-DMRG results are shown in Figure \ref{fig:Cr2}.
It is clear that \ce{Cr2} is much more challenging than \ce{C2},
since all the DMRG and p-DMRG errors for a given $M_0$  are larger than those for
\ce{C2} with the same $M_0$.
At the equilibrium geometry,
using $E_{d}^{(0)}$ ($\lambda=1$) in p-DMRG leads to relatively larger errors due to a near-intruder state,
while at $R$=1.50{\AA}, the second order energy is unphysically large.
Using the midpoint energy $\lambda=1/2$ as $E_0$ is a dramatic improvement
compared with both $\lambda=1$ and
$\lambda=0$ (the latter leads to an underestimation of the correlation energy).
With $M_0$ equal to 300 or 400, the p-DMRG($\lambda=1/2$) reaches
chemical accuracy, with the perturbation correction again providing
a large improvement of the variational energy. Thus, in the rest of this work, we
always use $\lambda=1/2$.

\subsection{\ce{Cr2} with (28e, 76o) orbital space}\label{sec:Cr2}

As a first example of a  larger calculation, we study the ground state energy
of \ce{Cr2} at $R$=1.68{\AA} with the cc-pVDZ-DK basis set\cite{doi:10.1063/1.1998907}.
Scalar relativistic effects were included through the spin-free X2C Hamiltonian\cite{liu2010ideas,SaueChemPhysChem2011,TheoChemAcc2012,spinSep2012}.
We used natural orbitals obtained from a CASSCF with a (12e,12o) active space in the
DMRG and p-DMRG calculations. The $1s$, $2s$ and $2p$ natural orbitals were not include in the (p)-DMRG calculations,
leading to an orbital space with (28e, 76o). The DMRG
and p-DMRG energies for \ce{Cr2}, as well as for the \ce{Cr} atom, are shown in Table \ref{table:Cr2_dz}.
As an empirical estimate, the extrapolation error bar in the variational DMRG is assigned as 1/5 of the difference
between the extrapolation energy and the energy with the largest $M=16000$\cite{chan_highly_2002}.

\begin{table}
  \caption{Energy ($E$+2099 in $E_h$) of \ce{Cr2} obtained with DMRG and p-DMRG
  in the cc-pVDZ-DK basis. The extrapolated DMRG energy of Cr atom is -1049.93254(4)$E_h$.}\label{table:Cr2_dz}

  \subcaption{Standard DMRG energy}
  \begin{tabular}{ccccccc}
    \hline
    \hline
    $M$ & 8000 & 10000 & 12000 & 14000 & 16000 & $\infty$ (extrapolated)\\
    \hline
    $E$ (default schedule) &-0.8957 & -0.8991 & -0.9024 & -0.9047 & -0.9061 & -0.9195$\pm$0.0027\\
    $E$ (reverse schedule) &-0.8980 & -0.9015 & -0.9040 & -0.9058 & -0.9071 & -0.9192$\pm$0.0024\\
    \hline
    \hline
  \end{tabular}

  \subcaption{p-DMRG energy: $E_2^{[i]}=\sum_{j=1}^{i}E_{2,j}$ represents the accumulated second-order perturbation energy for the sum of the first $i$ first-order MPS. $E_2^{(\infty)}$ represents the extrapolated energy for $M_1=\infty$. The final extrapolated p-DMRG energy with respect to $M_0$ is $E_\infty$=-2099.9201$E_h$.}
  \begin{tabular}{ccccc}
    \hline
    \hline
    $M_0$ & 1000 & 2000 & 3000 & 4000 \\
    \hline
    $E_{DMRG}^{(0)}$     & -0.8346  &  -0.8617  & -0.8743   &   -0.8818  \\
    \hline
    $E_2^{[1]}$ & -0.0607 &-0.0323 &-0.0196 &-0.0130  \\
    $E_2^{[2]}$ & -0.0652 &-0.0371 &-0.0243 &-0.0173  \\
    $E_2^{[3]}$ & -0.0671 &-0.0396 &-0.0268 &-0.0195  \\
    $E_2^{[4]}$ & -0.0682 &-0.0409 &-0.0282 &-0.0209  \\
    $E_2^{[5]}$ & -0.0690 &-0.0418 &-0.0293 &-0.0219  \\
    $E_2^{[\infty]}$ & -0.0734  &   -0.0492 &  -0.0386   &    -0.0323 \\
    \hline
    $E_{DMRG}^{(0)}+E_2^{[\infty]}$ & -0.9080 & -0.9109  &-0.9129   &   -0.9141 \\
    $\Delta_2/\Delta_0$${\;}^a$  & 0.141 & 0.157 & 0.157 & 0.157  \\
    \hline
    \hline
  \multicolumn{5}{l}{$^a$
  $\Delta_0 = E_{DMRG}^{(0)} - E_\infty$, $\Delta_2 = E_{DMRG}^{(0)}+E_2^{[\infty]} - E_\infty$.}
  \end{tabular}

\end{table}

As shown in Table \ref{table:Cr2_dz}(a), the standard variational DMRG energy converges very slowly
with respect to $M$. Even at $M=16000$, the variational DMRG energy is above
the extrapolated energy by about 10m$E_h$, while the DMRG energy at $M=8000$
is about 20m$E_h$ above.
Similarly, unlike in the p-DMRG calculation with (48e,42o), it is hard to converge $\ket{\Psi_1}$ with respect
to bond dimension using a single MPS. Thus, in this system we used the split ansatz \eqref{eq:split} to represent $\ket{\Psi_1}$.
We chose the bond dimension of each split MPS to be $M_1=7500$.
In Table \ref{table:Cr2_dz}(b), the accumulated second-order perturbation energies, $E_2^{[i]}=\sum_{j=1}^{i}E_{2,j}$ for the sum of the first $i$ first-order MPS, is shown for the first five terms in the split.
We also see slow convergence, for example,
at $M_0$=3000, adding an additional MPS in the sum only lowers the energy
by about 1m$E_h$ (after the second term in the sum).
In fact, we found that even after summing over 10 MPS (when $M_0$=3000), the change in $E_2$
for each subsequent MPS was as large as 0.3m$E_h$.
Thus, extrapolation is also needed to estimate a converged $E_2$.

\begin{figure}
\includegraphics[width=0.45\textwidth]{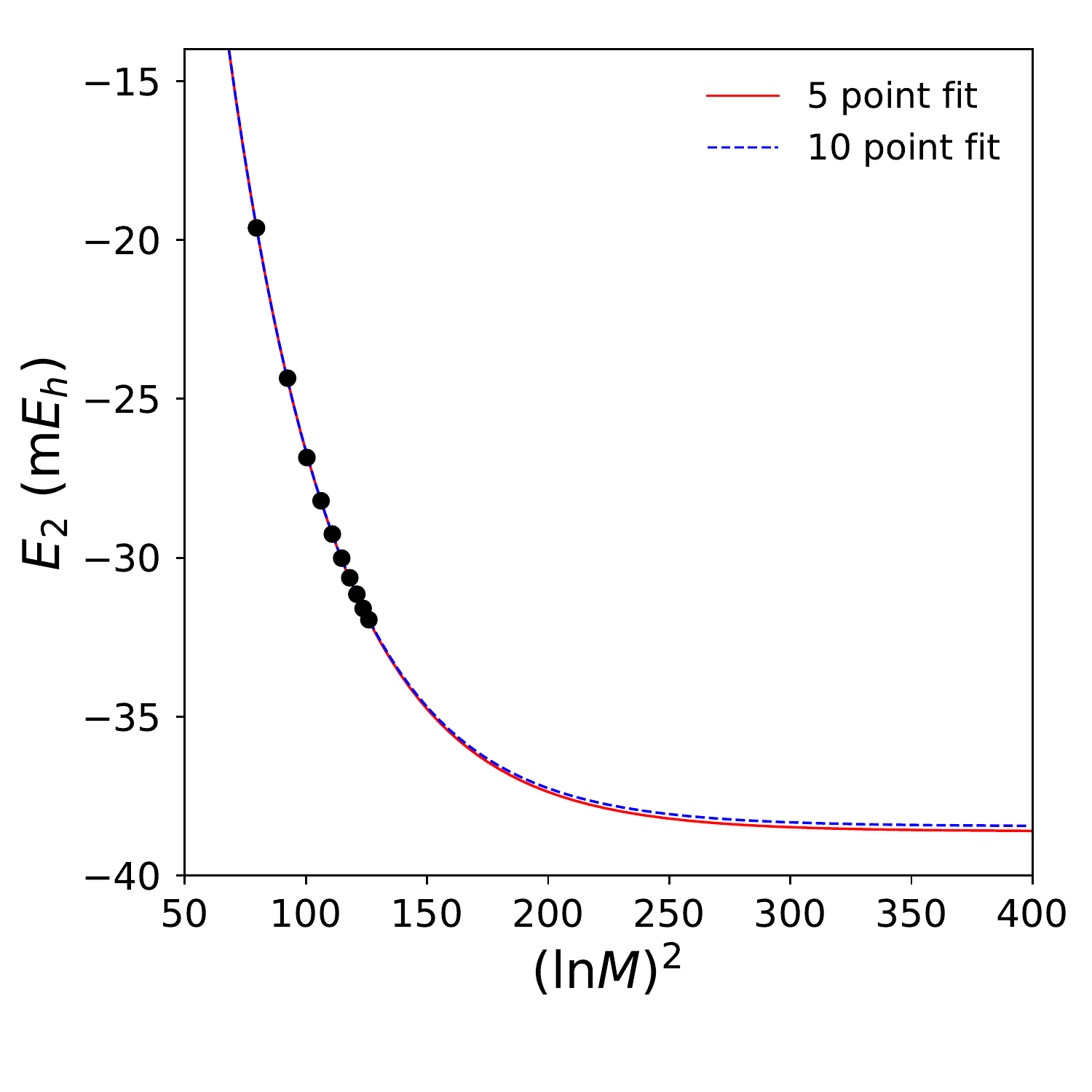}
\caption{The accumulated energies $E_2(M=NM_1)\triangleq E_2^{[N]}$ as a function of $(\ln M)^2$
and the fitted curves $E_2(M)=E_2^{[\infty]} + A e^{-\kappa(\ln M)^2}$
using the first 5 (red solid) and 10 (blue dashed) splitting functions for $M_0$=3000 and $M_1$=7500.}
\label{fig:Cr2_fitM1}
\end{figure}

To carry out the extrapolation, we used the linear relation between $\ln|\delta E|$ and $(\ln M)^2$ described in Refs. \cite{chan_highly_2002,doi:10.1063/1.1574318}.
Figure \ref{fig:Cr2_fitM1} shows the accumulated energies
$E_2(M=NM_1)\triangleq E_2^{[N]}$ as a function of $(\ln M)^2$
as well as the fitted curves $E_2(M)=E_2^{[\infty]} + A e^{-\kappa(\ln M)^2}$
using the first 5 (red solid) and 10 (blue dashed) points.
We see that using the first 5 points is sufficient to obtain a good extrapolation. The extrapolated $E_2^{[\infty]}$ from
5 points is -0.03861$E_h$, which
differs from that using 10 points (-0.03845$E_h$) by only 0.16m$E_h$.
Using such an extrapolation leads to substantial computational
savings. The full set of extrapolated results $E_2^{[\infty]}$ are listed in
Table \ref{table:Cr2_dz}(b).
It is notable that the p-DMRG energy at $M_0=1000$
with the first five basis functions, $E_{DMRG}^{[0]}+E_{2}^{[5]}$, is -0.9036$E_h$, which
is already close to the variational DMRG result
with $M_0=12000$. Using the extrapolated $E_2$, the p-DMRG energies are lower than the variational DMRG results.

To obtain a fully converged energy, we further need to extrapolate the variational bond dimension $M_0\to\infty$.
The need for two extrapolations are similar to the dual extrapolation in the original Heat-bath CI+PT\cite{holmes_heat-bath_2016}, where
one extrapolation is for the exact PT2 energy,
while the other is to extrapolate the CI energy to zero selection threshold.
To carry out this second extrapolation, we observe that the ratio $\Delta_2/\Delta_0$ where $\Delta_2 = E_{DMRG}^{(0)}+E_2^{[\infty]} - E_\infty$ and $\Delta_0 = E_{DMRG}^{(0)} - E_\infty$ is almost perfectly constant
for different $M_0$, as seen Table \ref{table:Cr2_dz}(b).
This relation allows us to estimate $E_\infty$.
The estimated $E_\infty$, using the largest three $M_0$, is -2099.9201$E_h$,
which is in agreement with the extrapolated variational DMRG results to within 1m$E_h$, and within the extrapolation error bars. Compared to the atomic energies, we obtain a binding energy at this geometry of 1.50 eV,
which is in fortuituously good agreement with the experimental value of of 1.47 eV\cite{casey_negative_1993}.
This demonstrates how, in practice, p-DMRG can be used as a cheaper alternative to variational DMRG to estimate
an exact ground state energy even in a fairly complicated system.

\subsection{Butadiene with (22e, 82o) active space}

The final system we consider is 1,3-butadiene. This system has been studied by many accurate methods including high-order coupled cluster theory\cite{watson_excited_2012} and $i$-FCIQMC\cite{daday_full_2012}. Benchmark energies have been reported using variational DMRG\cite{roberto}. We used the same basis ANO-L-VDZP[3s2p1d]/[2s1p] \cite{widmark_density_1990} as used in previous studies\cite{watson_excited_2012,daday_full_2012,roberto}.
All electrons except for a frozen 1s core were correlated, leading to
an orbital space with (22e, 82o). We used split-localized canonical orbitals for the p-DMRG calculations,
ordered by genetic ordering\cite{roberto}.
In the p-DMRG calculations, the first order MPS was split into five parts and each part had a bond dimension $M_1=3000$.
We used the same extrapolation procedures as used for \ce{Cr2} in the previous section.
The computed energies are shown in Table \ref{tab:butadiene}.
Due to the prohibitive computational cost, the extrapolated variational DMRG was not reported in Ref. \cite{roberto}.
However, it can be seen that
$E_{DMRG}^{(0)}+E_2^{[\infty]}$ for $M_0=2000$ is already lower
than the variational DMRG energy for $M=6000$.
Thus, we expect the exact ground state energy should be even lower.
Further using extrapolation for $M_0$, we obtain an estimated
exact energy of -155.557567$E_h$, which is lower than the $M_0=2000$ p-DMRG energy by only
0.25m$E_h$. Thus, we expect this extrapolated energy to be very close to the
exact ground state energy, and at least within the chemical accuracy.

\begin{table}
  \caption{Energy ($E$+155 in $E_h$) of butadiene with (22e,82o) active space.}
  \begin{tabular}{cccc}
    \hline
    \hline
    \multicolumn{4}{c}{DMRG-PT}\\
    \hline
    $M_0$ & $E_{DMRG}^{(0)}$ & $E_{DMRG}^{(0)}+E_2^{[\infty]}$  & $\Delta_2/\Delta_0$${\;}^a$ \\
    \hline
    500   & -0.552593 & -0.556038 & 0.308  \\
    1000  & -0.555438 & -0.556887 & 0.319   \\
    2000  & -0.556713 & -0.557318 & 0.292   \\
    $\infty$ & &-0.557567\\
    \hline
    $M=4000$ $^b$ & &-0.556874 &   \\
    $M=5000$ $^b$ & &-0.557050 &   \\
    $M=6000$ $^b$ & &-0.557178 &   \\
    \hline
    CCSD(T) $^c$&     &-0.555002  & \\
    CCSDT $^c$  &     &-0.555959  & \\
    $i$-FCIQMC $^d$  & &-0.5491(4)  & \\
    \hline
    \hline
  \multicolumn{4}{l}{$^a$
  $\Delta_0 = E_{DMRG}^{(0)} - E_\infty$, $\Delta_2 = E_{DMRG}^{(0)}+E_2^{[\infty]} - E_\infty$.}\\
  \multicolumn{4}{l}{$^b$ DMRG results from Ref. \citenum{roberto}.}\\
  \multicolumn{4}{l}{$^c$ Ref. \citenum{watson_excited_2012}.}\\
  \multicolumn{4}{l}{$^d$ Ref. \citenum{daday_full_2012}.}
  \end{tabular}
  \label{tab:butadiene}
\end{table}

\section{Conclusion}\label{conclusion}

In this work, we defined a p-DMRG method that uses perturbation theory within the DMRG framework to
efficiently target exact energies in large orbital spaces where not all orbitals are strongly correlated. 
Using a carefully defined zeroth order Hamiltonian, and with extrapolation procedures, 
we found that p-DMRG can indeed provide benchmark quality energies as accurate
as those obtained in far more expensive standard variational DMRG calculation.
Future work will be carried out to perform benchmark studies using p-DMRG for
the kinds of strongly correlated problems where there are a large number of intermediately correlated,
as well as strongly correlated orbitals, and which currently lie
beyond the capabilities of the practical variational DMRG calculations.


\begin{acknowledgement}
  This work was supported by the US National Science Foundation through CHE 1665333. Additional support
  was provided by OAC 1657286. ZL is supported by the Simons Collaboration on the Many-Electron Problem.
  GKC is a Simons Investigator in Physics.
\end{acknowledgement}

\providecommand{\latin}[1]{#1}
\providecommand*\mcitethebibliography{\thebibliography}
\csname @ifundefined\endcsname{endmcitethebibliography}
  {\let\endmcitethebibliography\endthebibliography}{}


\begin{mcitethebibliography}{59}
\providecommand*\natexlab[1]{#1}
\providecommand*\mciteSetBstSublistMode[1]{}
\providecommand*\mciteSetBstMaxWidthForm[2]{}
\providecommand*\mciteBstWouldAddEndPuncttrue
  {\def\EndOfBibitem{\unskip.}}
\providecommand*\mciteBstWouldAddEndPunctfalse
  {\let\EndOfBibitem\relax}
\providecommand*\mciteSetBstMidEndSepPunct[3]{}
\providecommand*\mciteSetBstSublistLabelBeginEnd[3]{}
\providecommand*\EndOfBibitem{}
\mciteSetBstSublistMode{f}
\mciteSetBstMaxWidthForm{subitem}{(\alph{mcitesubitemcount})}
\mciteSetBstSublistLabelBeginEnd
  {\mcitemaxwidthsubitemform\space}
  {\relax}
  {\relax}

\bibitem[White(1992)]{white_density_1992}
White,~S.~R. Density matrix formulation for quantum renormalization groups.
  \emph{Phys. Rev. Lett.} \textbf{1992}, \emph{69}, 2863--2866\relax
\mciteBstWouldAddEndPuncttrue
\mciteSetBstMidEndSepPunct{\mcitedefaultmidpunct}
{\mcitedefaultendpunct}{\mcitedefaultseppunct}\relax
\EndOfBibitem
\bibitem[White(1993)]{white_density-matrix_1993}
White,~S.~R. Density-matrix algorithms for quantum renormalization groups.
  \emph{Phys. Rev. B} \textbf{1993}, \emph{48}, 10345--10356\relax
\mciteBstWouldAddEndPuncttrue
\mciteSetBstMidEndSepPunct{\mcitedefaultmidpunct}
{\mcitedefaultendpunct}{\mcitedefaultseppunct}\relax
\EndOfBibitem
\bibitem[White and Martin(1999)White, and Martin]{white_ab_1999}
White,~S.~R.; Martin,~R.~L. Ab initio quantum chemistry using the density
  matrix renormalization group. \emph{J. Chem. Phys.} \textbf{1999},
  \emph{110}, 4127--4130\relax
\mciteBstWouldAddEndPuncttrue
\mciteSetBstMidEndSepPunct{\mcitedefaultmidpunct}
{\mcitedefaultendpunct}{\mcitedefaultseppunct}\relax
\EndOfBibitem
\bibitem[Mitrushenkov \latin{et~al.}(2001)Mitrushenkov, Fano, Ortolani,
  Linguerri, and Palmieri]{mitrushenkov2001quantum}
Mitrushenkov,~A.~O.; Fano,~G.; Ortolani,~F.; Linguerri,~R.; Palmieri,~P.
  {Quantum chemistry using the density matrix renormalization group}. \emph{J.
  Chem. Phys.} \textbf{2001}, \emph{115}, 6815--6821\relax
\mciteBstWouldAddEndPuncttrue
\mciteSetBstMidEndSepPunct{\mcitedefaultmidpunct}
{\mcitedefaultendpunct}{\mcitedefaultseppunct}\relax
\EndOfBibitem
\bibitem[Chan and Head-Gordon(2002)Chan, and Head-Gordon]{chan_highly_2002}
Chan,~G. K.-L.; Head-Gordon,~M. Highly correlated calculations with a
  polynomial cost algorithm: {A} study of the density matrix renormalization
  group. \emph{J. Chem. Phys.} \textbf{2002}, \emph{116}, 4462--4476\relax
\mciteBstWouldAddEndPuncttrue
\mciteSetBstMidEndSepPunct{\mcitedefaultmidpunct}
{\mcitedefaultendpunct}{\mcitedefaultseppunct}\relax
\EndOfBibitem
\bibitem[Legeza \latin{et~al.}(2003)Legeza, R{\"o}der, and
  Hess]{legeza2003controlling}
Legeza,~{\"O}.; R{\"o}der,~J.; Hess,~B.~A. {Controlling the accuracy of the
  density-matrix renormalization-group method: The dynamical block state
  selection approach}. \emph{Phys. Rev. B} \textbf{2003}, \emph{67},
  125114\relax
\mciteBstWouldAddEndPuncttrue
\mciteSetBstMidEndSepPunct{\mcitedefaultmidpunct}
{\mcitedefaultendpunct}{\mcitedefaultseppunct}\relax
\EndOfBibitem
\bibitem[Sharma and Chan(2012)Sharma, and Chan]{sharma_spin-adapted_2012}
Sharma,~S.; Chan,~G. K.-L. Spin-adapted density matrix renormalization group
  algorithms for quantum chemistry. \emph{J. Chem. Phys.} \textbf{2012},
  \emph{136}, 124121\relax
\mciteBstWouldAddEndPuncttrue
\mciteSetBstMidEndSepPunct{\mcitedefaultmidpunct}
{\mcitedefaultendpunct}{\mcitedefaultseppunct}\relax
\EndOfBibitem
\bibitem[Olivares-Amaya \latin{et~al.}(2015)Olivares-Amaya, Hu, Nakatani,
  Sharma, Yang, and Chan]{roberto}
Olivares-Amaya,~R.; Hu,~W.; Nakatani,~N.; Sharma,~S.; Yang,~J.; Chan,~G. K.-L.
  The ab-initio density matrix renormalization group in practice. \emph{J.
  Chem. Phys.} \textbf{2015}, \emph{142}, 034102\relax
\mciteBstWouldAddEndPuncttrue
\mciteSetBstMidEndSepPunct{\mcitedefaultmidpunct}
{\mcitedefaultendpunct}{\mcitedefaultseppunct}\relax
\EndOfBibitem
\bibitem[Keller \latin{et~al.}(2015)Keller, Dolfi, Troyer, and
  Reiher]{keller_efficient_2015}
Keller,~S.; Dolfi,~M.; Troyer,~M.; Reiher,~M. An efficient matrix product
  operator representation of the quantum chemical Hamiltonian. \emph{J. Chem.
  Phys.} \textbf{2015}, \emph{143}, 244118\relax
\mciteBstWouldAddEndPuncttrue
\mciteSetBstMidEndSepPunct{\mcitedefaultmidpunct}
{\mcitedefaultendpunct}{\mcitedefaultseppunct}\relax
\EndOfBibitem
\bibitem[Yanai \latin{et~al.}(2015)Yanai, Kurashige, Mizukami, Chalupsk\'{y},
  Lan, and Saitow]{yanai_density_2015}
Yanai,~T.; Kurashige,~Y.; Mizukami,~W.; Chalupsk\'{y},~J.; Lan,~T.~N.;
  Saitow,~M. Density matrix renormalization group for ab initio {Calculations}
  and associated dynamic correlation methods: {A} review of theory and
  applications. \emph{Int. J. Quantum Chem.} \textbf{2015}, \emph{115},
  283--299\relax
\mciteBstWouldAddEndPuncttrue
\mciteSetBstMidEndSepPunct{\mcitedefaultmidpunct}
{\mcitedefaultendpunct}{\mcitedefaultseppunct}\relax
\EndOfBibitem
\bibitem[Chan \latin{et~al.}(2016)Chan, Keselman, Nakatani, Li, and
  White]{chan_matrix_2016}
Chan,~G. K.-L.; Keselman,~A.; Nakatani,~N.; Li,~Z.; White,~S.~R. Matrix product
  operators, matrix product states, and ab initio density matrix
  renormalization group algorithms. \emph{J. Chem. Phys.} \textbf{2016},
  \emph{145}, 014102\relax
\mciteBstWouldAddEndPuncttrue
\mciteSetBstMidEndSepPunct{\mcitedefaultmidpunct}
{\mcitedefaultendpunct}{\mcitedefaultseppunct}\relax
\EndOfBibitem
\bibitem[Andersson \latin{et~al.}(1990)Andersson, Malmqvist, Roos, Sadlej, and
  Wolinski]{andersson_second-order_1990}
Andersson,~K.; Malmqvist,~P.~A.; Roos,~B.~O.; Sadlej,~A.~J.; Wolinski,~K.
  Second-order perturbation theory with a {CASSCF} reference function. \emph{J.
  Phys. Chem.} \textbf{1990}, \emph{94}, 5483--5488\relax
\mciteBstWouldAddEndPuncttrue
\mciteSetBstMidEndSepPunct{\mcitedefaultmidpunct}
{\mcitedefaultendpunct}{\mcitedefaultseppunct}\relax
\EndOfBibitem
\bibitem[Roos \latin{et~al.}(1996)Roos, Andersson, FÃ¼lscher, Malmqvist,
  Serrano-AndrÃ©s, Pierloot, and MerchÃ¡n]{roos_multiconfigurational_1996}
Roos,~B.~O.; Andersson,~K.; FÃ¼lscher,~M.~P.; Malmqvist,~P.-Ã.;
  Serrano-AndrÃ©s,~L.; Pierloot,~K.; MerchÃ¡n,~M. In \emph{Advances in
  {Chemical} {Physics}}; Prigogine,~I., Rice,~S.~A., Eds.; John Wiley \& Sons,
  Inc., 1996; pp 219--331\relax
\mciteBstWouldAddEndPuncttrue
\mciteSetBstMidEndSepPunct{\mcitedefaultmidpunct}
{\mcitedefaultendpunct}{\mcitedefaultseppunct}\relax
\EndOfBibitem
\bibitem[Angeli \latin{et~al.}(2001)Angeli, Cimiraglia, Evangelisti, Leininger,
  and Malrieu]{angeli_introduction_2001}
Angeli,~C.; Cimiraglia,~R.; Evangelisti,~S.; Leininger,~T.; Malrieu,~J.-P.
  Introduction of n-electron valence states for multireference perturbation
  theory. \emph{J. Chem. Phys.} \textbf{2001}, \emph{114}, 10252--10264\relax
\mciteBstWouldAddEndPuncttrue
\mciteSetBstMidEndSepPunct{\mcitedefaultmidpunct}
{\mcitedefaultendpunct}{\mcitedefaultseppunct}\relax
\EndOfBibitem
\bibitem[Angeli \latin{et~al.}(2001)Angeli, Cimiraglia, and
  Malrieu]{angeli_n-electron_2001}
Angeli,~C.; Cimiraglia,~R.; Malrieu,~J.-P. N-electron valence state
  perturbation theory: a fast implementation of the strongly contracted
  variant. \emph{Chem. Phys. Lett.} \textbf{2001}, \emph{350}, 297--305\relax
\mciteBstWouldAddEndPuncttrue
\mciteSetBstMidEndSepPunct{\mcitedefaultmidpunct}
{\mcitedefaultendpunct}{\mcitedefaultseppunct}\relax
\EndOfBibitem
\bibitem[Angeli \latin{et~al.}(2002)Angeli, Cimiraglia, and
  Malrieu]{angeli_n-electron_2002}
Angeli,~C.; Cimiraglia,~R.; Malrieu,~J.-P. n-electron valence state
  perturbation theory: {A} spinless formulation and an efficient implementation
  of the strongly contracted and of the partially contracted variants. \emph{J.
  Chem. Phys.} \textbf{2002}, \emph{117}, 9138--9153\relax
\mciteBstWouldAddEndPuncttrue
\mciteSetBstMidEndSepPunct{\mcitedefaultmidpunct}
{\mcitedefaultendpunct}{\mcitedefaultseppunct}\relax
\EndOfBibitem
\bibitem[Kurashige and Yanai(2011)Kurashige, and
  Yanai]{kurashige_second-order_2011}
Kurashige,~Y.; Yanai,~T. Second-order perturbation theory with a density matrix
  renormalization group self-consistent field reference function: {Theory} and
  application to the study of chromium dimer. \emph{J. Chem. Phys.}
  \textbf{2011}, \emph{135}, 094104\relax
\mciteBstWouldAddEndPuncttrue
\mciteSetBstMidEndSepPunct{\mcitedefaultmidpunct}
{\mcitedefaultendpunct}{\mcitedefaultseppunct}\relax
\EndOfBibitem
\bibitem[Sharma and Chan(2014)Sharma, and Chan]{sharma_communication:_2014}
Sharma,~S.; Chan,~G. K.-L. Communication: {A} flexible multi-reference
  perturbation theory by minimizing the {Hylleraas} functional with matrix
  product states. \emph{J. Chem. Phys.} \textbf{2014}, \emph{141}, 111101\relax
\mciteBstWouldAddEndPuncttrue
\mciteSetBstMidEndSepPunct{\mcitedefaultmidpunct}
{\mcitedefaultendpunct}{\mcitedefaultseppunct}\relax
\EndOfBibitem
\bibitem[Guo \latin{et~al.}(2016)Guo, Watson, Hu, Sun, and
  Chan]{guo_n-electron_2016}
Guo,~S.; Watson,~M.~A.; Hu,~W.; Sun,~Q.; Chan,~G. K.-L. N-{Electron} {Valence}
  {State} {Perturbation} {Theory} {Based} on a {Density} {Matrix}
  {Renormalization} {Group} {Reference} {Function}, with {Applications} to the
  {Chromium} {Dimer} and a {Trimer} {Model} of {Poly}(p-{Phenylenevinylene}).
  \emph{J. Chem. Theory Comput.} \textbf{2016}, \emph{12}, 1583--1591\relax
\mciteBstWouldAddEndPuncttrue
\mciteSetBstMidEndSepPunct{\mcitedefaultmidpunct}
{\mcitedefaultendpunct}{\mcitedefaultseppunct}\relax
\EndOfBibitem
\bibitem[Sokolov and Chan(2016)Sokolov, and Chan]{Alex_t_nevpt}
Sokolov,~A.~Y.; Chan,~G. K.-L. A time-dependent formulation of multi-reference
  perturbation theory. \emph{J. Chem. Phys.} \textbf{2016}, \emph{144},
  064102\relax
\mciteBstWouldAddEndPuncttrue
\mciteSetBstMidEndSepPunct{\mcitedefaultmidpunct}
{\mcitedefaultendpunct}{\mcitedefaultseppunct}\relax
\EndOfBibitem
\bibitem[Freitag \latin{et~al.}(2017)Freitag, Knecht, Angeli, and
  Reiher]{Reigher_dmrg_nevptpt}
Freitag,~L.; Knecht,~S.; Angeli,~C.; Reiher,~M. Multireference Perturbation
  Theory with Cholesky Decomposition for the Density Matrix Renormalization
  Group. \emph{J. Chem. Theory Comput.} \textbf{2017}, \emph{13}, 451--459,
  PMID: 28094988\relax
\mciteBstWouldAddEndPuncttrue
\mciteSetBstMidEndSepPunct{\mcitedefaultmidpunct}
{\mcitedefaultendpunct}{\mcitedefaultseppunct}\relax
\EndOfBibitem
\bibitem[Sokolov \latin{et~al.}(2017)Sokolov, Guo, Ronca, and
  Chan]{Alex_t_dmrg_pt}
Sokolov,~A.~Y.; Guo,~S.; Ronca,~E.; Chan,~G. K.-L. Time-dependent N-electron
  valence perturbation theory with matrix product state reference wavefunctions
  for large active spaces and basis sets: Applications to the chromium dimer
  and all-trans polyenes. \emph{J. Chem. Phys.} \textbf{2017}, \emph{146},
  244102\relax
\mciteBstWouldAddEndPuncttrue
\mciteSetBstMidEndSepPunct{\mcitedefaultmidpunct}
{\mcitedefaultendpunct}{\mcitedefaultseppunct}\relax
\EndOfBibitem
\bibitem[Nakatani and Guo(2017)Nakatani, and Guo]{Naoki_caspt2}
Nakatani,~N.; Guo,~S. Density matrix renormalization group (DMRG) method as a
  common tool for large active-space CASSCF/CASPT2 calculations. \emph{J. Chem.
  Phys.} \textbf{2017}, \emph{146}, 094102\relax
\mciteBstWouldAddEndPuncttrue
\mciteSetBstMidEndSepPunct{\mcitedefaultmidpunct}
{\mcitedefaultendpunct}{\mcitedefaultseppunct}\relax
\EndOfBibitem
\bibitem[Huron \latin{et~al.}(1973)Huron, Malrieu, and
  Rancurel]{huron1973iterative}
Huron,~B.; Malrieu,~J.; Rancurel,~P. Iterative perturbation calculations of
  ground and excited state energies from multiconfigurational zeroth-order
  wavefunctions. \emph{J. Chem. Phys.} \textbf{1973}, \emph{58},
  5745--5759\relax
\mciteBstWouldAddEndPuncttrue
\mciteSetBstMidEndSepPunct{\mcitedefaultmidpunct}
{\mcitedefaultendpunct}{\mcitedefaultseppunct}\relax
\EndOfBibitem
\bibitem[Buenker and Peyerimhoff(1974)Buenker, and
  Peyerimhoff]{buenker1974individualized}
Buenker,~R.~J.; Peyerimhoff,~S.~D. Individualized configuration selection in CI
  calculations with subsequent energy extrapolation. \emph{Theor. Chem. Acc.}
  \textbf{1974}, \emph{35}, 33--58\relax
\mciteBstWouldAddEndPuncttrue
\mciteSetBstMidEndSepPunct{\mcitedefaultmidpunct}
{\mcitedefaultendpunct}{\mcitedefaultseppunct}\relax
\EndOfBibitem
\bibitem[Harrison(1991)]{harrison1991approximating}
Harrison,~R.~J. Approximating full configuration interaction with selected
  configuration interaction and perturbation theory. \emph{J. Chem. Phys.}
  \textbf{1991}, \emph{94}, 5021--5031\relax
\mciteBstWouldAddEndPuncttrue
\mciteSetBstMidEndSepPunct{\mcitedefaultmidpunct}
{\mcitedefaultendpunct}{\mcitedefaultseppunct}\relax
\EndOfBibitem
\bibitem[Schriber and Evangelista(2016)Schriber, and
  Evangelista]{schriber2016communication}
Schriber,~J.~B.; Evangelista,~F.~A. Communication: An adaptive configuration
  interaction approach for strongly correlated electrons with tunable accuracy.
  \emph{J. Chem. Phys.} \textbf{2016}, \emph{144}, 161106\relax
\mciteBstWouldAddEndPuncttrue
\mciteSetBstMidEndSepPunct{\mcitedefaultmidpunct}
{\mcitedefaultendpunct}{\mcitedefaultseppunct}\relax
\EndOfBibitem
\bibitem[Tubman \latin{et~al.}(2016)Tubman, Lee, Takeshita, Head-Gordon, and
  Whaley]{tubman2016deterministic}
Tubman,~N.~M.; Lee,~J.; Takeshita,~T.~Y.; Head-Gordon,~M.; Whaley,~K.~B. A
  deterministic alternative to the full configuration interaction quantum Monte
  Carlo method. \emph{J. Chem. Phys.} \textbf{2016}, \emph{145}, 044112\relax
\mciteBstWouldAddEndPuncttrue
\mciteSetBstMidEndSepPunct{\mcitedefaultmidpunct}
{\mcitedefaultendpunct}{\mcitedefaultseppunct}\relax
\EndOfBibitem
\bibitem[Liu and Hoffmann(2016)Liu, and Hoffmann]{liu2016ici}
Liu,~W.; Hoffmann,~M.~R. iCI: Iterative CI toward full CI. \emph{J. Chem.
  Theory Comput.} \textbf{2016}, \emph{12}, 1169--1178\relax
\mciteBstWouldAddEndPuncttrue
\mciteSetBstMidEndSepPunct{\mcitedefaultmidpunct}
{\mcitedefaultendpunct}{\mcitedefaultseppunct}\relax
\EndOfBibitem
\bibitem[Holmes \latin{et~al.}(2016)Holmes, Tubman, and
  Umrigar]{holmes_heat-bath_2016}
Holmes,~A.~A.; Tubman,~N.~M.; Umrigar,~C.~J. Heat-{Bath} {Configuration}
  {Interaction}: {An} {Efficient} {Selected} {Configuration} {Interaction}
  {Algorithm} {Inspired} by {Heat}-{Bath} {Sampling}. \emph{J. Chem. Theory
  Comput.} \textbf{2016}, \emph{12}, 3674--3680\relax
\mciteBstWouldAddEndPuncttrue
\mciteSetBstMidEndSepPunct{\mcitedefaultmidpunct}
{\mcitedefaultendpunct}{\mcitedefaultseppunct}\relax
\EndOfBibitem
\bibitem[Sharma \latin{et~al.}(2017)Sharma, Holmes, Jeanmairet, Alavi, and
  Umrigar]{sharma2017semistochastic}
Sharma,~S.; Holmes,~A.~A.; Jeanmairet,~G.; Alavi,~A.; Umrigar,~C.~J.
  Semistochastic Heat-bath Configuration Interaction method: selected
  configuration interaction with semistochastic perturbation theory. \emph{J.
  Chem. Theory Comput.} \textbf{2017}, \emph{13}, 1595--1604\relax
\mciteBstWouldAddEndPuncttrue
\mciteSetBstMidEndSepPunct{\mcitedefaultmidpunct}
{\mcitedefaultendpunct}{\mcitedefaultseppunct}\relax
\EndOfBibitem
\bibitem[Garniron \latin{et~al.}(2017)Garniron, Scemama, Loos, and
  Caffarel]{garniron2017hybrid}
Garniron,~Y.; Scemama,~A.; Loos,~P.-F.; Caffarel,~M. Hybrid
  stochastic-deterministic calculation of the second-order perturbative
  contribution of multireference perturbation theory. \emph{J. Chem. Phys.}
  \textbf{2017}, \emph{147}, 034101\relax
\mciteBstWouldAddEndPuncttrue
\mciteSetBstMidEndSepPunct{\mcitedefaultmidpunct}
{\mcitedefaultendpunct}{\mcitedefaultseppunct}\relax
\EndOfBibitem
\bibitem[Foster and Boys(1960)Foster, and Boys]{foster1960canonical}
Foster,~J.; Boys,~S. Canonical configurational interaction procedure.
  \emph{Rev. Mod. Phys.} \textbf{1960}, \emph{32}, 300\relax
\mciteBstWouldAddEndPuncttrue
\mciteSetBstMidEndSepPunct{\mcitedefaultmidpunct}
{\mcitedefaultendpunct}{\mcitedefaultseppunct}\relax
\EndOfBibitem
\bibitem[Bender and Davidson(1969)Bender, and Davidson]{bender1969studies}
Bender,~C.~F.; Davidson,~E.~R. Studies in configuration interaction: The
  first-row diatomic hydrides. \emph{Phys. Rev.} \textbf{1969}, \emph{183},
  23\relax
\mciteBstWouldAddEndPuncttrue
\mciteSetBstMidEndSepPunct{\mcitedefaultmidpunct}
{\mcitedefaultendpunct}{\mcitedefaultseppunct}\relax
\EndOfBibitem
\bibitem[Hachmann \latin{et~al.}(2006)Hachmann, Cardoen, and
  Chan]{Hachmann_Hchain}
Hachmann,~J.; Cardoen,~W.; Chan,~G. K.-L. Multireference correlation in long
  molecules with the quadratic scaling density matrix renormalization group.
  \emph{J. Chem. Phys.} \textbf{2006}, \emph{125}, 144101\relax
\mciteBstWouldAddEndPuncttrue
\mciteSetBstMidEndSepPunct{\mcitedefaultmidpunct}
{\mcitedefaultendpunct}{\mcitedefaultseppunct}\relax
\EndOfBibitem
\bibitem[Hylleraas(1930)]{Hylleraas_functional}
Hylleraas,~E.~A. {\"U}ber den Grundterm der Zweielektronenprobleme von Hâ,
  He, Li+, Be++ usw. \emph{Z. Phys.} \textbf{1930}, \emph{65}, 209--225\relax
\mciteBstWouldAddEndPuncttrue
\mciteSetBstMidEndSepPunct{\mcitedefaultmidpunct}
{\mcitedefaultendpunct}{\mcitedefaultseppunct}\relax
\EndOfBibitem
\bibitem[Ren \latin{et~al.}(2016)Ren, Yi, and Shuai]{ren_inner_2016}
Ren,~J.; Yi,~Y.; Shuai,~Z. Inner {Space} {Perturbation} {Theory} in {Matrix}
  {Product} {States}: {Replacing} {Expensive} {Iterative} {Diagonalization}.
  \emph{J. Chem. Theory Comput.} \textbf{2016}, \emph{12}, 4871--4878\relax
\mciteBstWouldAddEndPuncttrue
\mciteSetBstMidEndSepPunct{\mcitedefaultmidpunct}
{\mcitedefaultendpunct}{\mcitedefaultseppunct}\relax
\EndOfBibitem
\bibitem[Schollw{\"o}ck(2011)]{schollwock_density-matrix_2011}
Schollw{\"o}ck,~U. The density-matrix renormalization group in the age of
  matrix product states. \emph{Ann. Phys.} \textbf{2011}, \emph{326},
  96--192\relax
\mciteBstWouldAddEndPuncttrue
\mciteSetBstMidEndSepPunct{\mcitedefaultmidpunct}
{\mcitedefaultendpunct}{\mcitedefaultseppunct}\relax
\EndOfBibitem
\bibitem[Keller and Reiher(2016)Keller, and Reiher]{keller2016spin}
Keller,~S.; Reiher,~M. {Spin-adapted matrix product states and operators}.
  \emph{J. Chem. Phys.} \textbf{2016}, \emph{144}, 134101\relax
\mciteBstWouldAddEndPuncttrue
\mciteSetBstMidEndSepPunct{\mcitedefaultmidpunct}
{\mcitedefaultendpunct}{\mcitedefaultseppunct}\relax
\EndOfBibitem
\bibitem[Xiang(1996)]{xiang1996density}
Xiang,~T. {Density-matrix renormalization-group method in momentum space}.
  \emph{Phys Rev. B} \textbf{1996}, \emph{53}, R10445\relax
\mciteBstWouldAddEndPuncttrue
\mciteSetBstMidEndSepPunct{\mcitedefaultmidpunct}
{\mcitedefaultendpunct}{\mcitedefaultseppunct}\relax
\EndOfBibitem
\bibitem[Li and Chan(2017)Li, and Chan]{li2017spmps}
Li,~Z.; Chan,~G. K.-L. Spin-Projected Matrix Product States: Versatile Tool for
  Strongly Correlated Systems. \emph{J. Chem. Theory Comput.} \textbf{2017},
  \emph{13}, 2681--2695\relax
\mciteBstWouldAddEndPuncttrue
\mciteSetBstMidEndSepPunct{\mcitedefaultmidpunct}
{\mcitedefaultendpunct}{\mcitedefaultseppunct}\relax
\EndOfBibitem
\bibitem[Surj{\'a}n and Szabados(2000)Surj{\'a}n, and
  Szabados]{surjan2000optimized}
Surj{\'a}n,~P.; Szabados,~A. Optimized partitioning in perturbation theory:
  Comparison to related approaches. \emph{J. Chem. Phys.} \textbf{2000},
  \emph{112}, 4438--4446\relax
\mciteBstWouldAddEndPuncttrue
\mciteSetBstMidEndSepPunct{\mcitedefaultmidpunct}
{\mcitedefaultendpunct}{\mcitedefaultseppunct}\relax
\EndOfBibitem
\bibitem[Hehre \latin{et~al.}(1969)Hehre, Stewart, and
  Pople]{doi:10.1063/1.1672392}
Hehre,~W.~J.; Stewart,~R.~F.; Pople,~J.~A. Self-Consistent Molecular-Orbital
  Methods. I. Use of Gaussian Expansions of Slater-Type Atomic Orbitals.
  \emph{J. Chem. Phys.} \textbf{1969}, \emph{51}, 2657--2664\relax
\mciteBstWouldAddEndPuncttrue
\mciteSetBstMidEndSepPunct{\mcitedefaultmidpunct}
{\mcitedefaultendpunct}{\mcitedefaultseppunct}\relax
\EndOfBibitem
\bibitem[Jr.(1970)]{doi:10.1063/1.1674408}
Jr.,~T. H.~D. Gaussian Basis Functions for Use in Molecular Calculations. I.
  Contraction of (9s5p) Atomic Basis Sets for the FirstâRow Atoms. \emph{J.
  Chem. Phys.} \textbf{1970}, \emph{53}, 2823--2833\relax
\mciteBstWouldAddEndPuncttrue
\mciteSetBstMidEndSepPunct{\mcitedefaultmidpunct}
{\mcitedefaultendpunct}{\mcitedefaultseppunct}\relax
\EndOfBibitem
\bibitem[Huang \latin{et~al.}()Huang, Liao, Liu, Xie, Xie, Zhao, Chen, and
  Xiang]{split_mps}
Huang,~R.-Z.; Liao,~H.-J.; Liu,~Z.-Y.; Xie,~H.-D.; Xie,~Z.-Y.; Zhao,~H.-H.;
  Chen,~J.; Xiang,~T. A generalized Lanczos method for systematic optimization
  of tensor network states. \emph{arXiv:1611.09574} \relax
\mciteBstWouldAddEndPunctfalse
\mciteSetBstMidEndSepPunct{\mcitedefaultmidpunct}
{}{\mcitedefaultseppunct}\relax
\EndOfBibitem
\bibitem[Kendall \latin{et~al.}(1992)Kendall, Jr., and
  Harrison]{doi:10.1063/1.462569}
Kendall,~R.~A.; Jr.,~T. H.~D.; Harrison,~R.~J. Electron affinities of the
  first-row atoms revisited. Systematic basis sets and wave functions. \emph{J.
  Chem. Phys.} \textbf{1992}, \emph{96}, 6796--6806\relax
\mciteBstWouldAddEndPuncttrue
\mciteSetBstMidEndSepPunct{\mcitedefaultmidpunct}
{\mcitedefaultendpunct}{\mcitedefaultseppunct}\relax
\EndOfBibitem
\bibitem[Bondybey and English(1983)Bondybey, and
  English]{bondybey_electronic_1983}
Bondybey,~V.~E.; English,~J.~H. Electronic structure and vibrational frequency
  of {Cr}2. \emph{Chem. Phys. Lett.} \textbf{1983}, \emph{94}, 443--447\relax
\mciteBstWouldAddEndPuncttrue
\mciteSetBstMidEndSepPunct{\mcitedefaultmidpunct}
{\mcitedefaultendpunct}{\mcitedefaultseppunct}\relax
\EndOfBibitem
\bibitem[Sch{\"a}fer \latin{et~al.}(1992)Sch{\"a}fer, Horn, and
  Ahlrichs]{doi:10.1063/1.463096}
Sch{\"a}fer,~A.; Horn,~H.; Ahlrichs,~R. Fully optimized contracted Gaussian
  basis sets for atoms Li to Kr. \emph{J. Chem. Phys.} \textbf{1992},
  \emph{97}, 2571--2577\relax
\mciteBstWouldAddEndPuncttrue
\mciteSetBstMidEndSepPunct{\mcitedefaultmidpunct}
{\mcitedefaultendpunct}{\mcitedefaultseppunct}\relax
\EndOfBibitem
\bibitem[Balabanov and Peterson(2005)Balabanov, and
  Peterson]{doi:10.1063/1.1998907}
Balabanov,~N.~B.; Peterson,~K.~A. Systematically convergent basis sets for
  transition metals. I. All-electron correlation consistent basis sets for the
  3d elements Sc-Zn. \emph{J. Chem. Phys.} \textbf{2005}, \emph{123},
  064107\relax
\mciteBstWouldAddEndPuncttrue
\mciteSetBstMidEndSepPunct{\mcitedefaultmidpunct}
{\mcitedefaultendpunct}{\mcitedefaultseppunct}\relax
\EndOfBibitem
\bibitem[Liu(2010)]{liu2010ideas}
Liu,~W. Ideas of relativistic quantum chemistry. \emph{Mol. Phys.}
  \textbf{2010}, \emph{108}, 1679--1706\relax
\mciteBstWouldAddEndPuncttrue
\mciteSetBstMidEndSepPunct{\mcitedefaultmidpunct}
{\mcitedefaultendpunct}{\mcitedefaultseppunct}\relax
\EndOfBibitem
\bibitem[Saue(2011)]{SaueChemPhysChem2011}
Saue,~T. Relativistic Hamiltonians for Chemistry: A Primer. \emph{ChemPhysChem}
  \textbf{2011}, \emph{12}, 3077--3094\relax
\mciteBstWouldAddEndPuncttrue
\mciteSetBstMidEndSepPunct{\mcitedefaultmidpunct}
{\mcitedefaultendpunct}{\mcitedefaultseppunct}\relax
\EndOfBibitem
\bibitem[Peng and Reiher(2012)Peng, and Reiher]{TheoChemAcc2012}
Peng,~D.; Reiher,~M. Exact decoupling of the relativistic Fock operator.
  \emph{Theor. Chem. Acc.} \textbf{2012}, \emph{131}, 1081\relax
\mciteBstWouldAddEndPuncttrue
\mciteSetBstMidEndSepPunct{\mcitedefaultmidpunct}
{\mcitedefaultendpunct}{\mcitedefaultseppunct}\relax
\EndOfBibitem
\bibitem[Li \latin{et~al.}(2012)Li, Xiao, and Liu]{spinSep2012}
Li,~Z.; Xiao,~Y.; Liu,~W. On the spin separation of algebraic two-component
  relativistic Hamiltonians. \emph{J. Chem. Phys.} \textbf{2012}, \emph{137},
  154114\relax
\mciteBstWouldAddEndPuncttrue
\mciteSetBstMidEndSepPunct{\mcitedefaultmidpunct}
{\mcitedefaultendpunct}{\mcitedefaultseppunct}\relax
\EndOfBibitem
\bibitem[Chan and Head-Gordon(2003)Chan, and
  Head-Gordon]{doi:10.1063/1.1574318}
Chan,~G. K.-L.; Head-Gordon,~M. Exact solution (within a triple-zeta, double
  polarization basis set) of the electronic SchrÃ¶dinger equation for water.
  \emph{J. Chem. Phys.} \textbf{2003}, \emph{118}, 8551--8554\relax
\mciteBstWouldAddEndPuncttrue
\mciteSetBstMidEndSepPunct{\mcitedefaultmidpunct}
{\mcitedefaultendpunct}{\mcitedefaultseppunct}\relax
\EndOfBibitem
\bibitem[Casey and Leopold(1993)Casey, and Leopold]{casey_negative_1993}
Casey,~S.~M.; Leopold,~D.~G. Negative ion photoelectron spectroscopy of
  chromium dimer. \emph{J. Phys. Chem.} \textbf{1993}, \emph{97},
  816--830\relax
\mciteBstWouldAddEndPuncttrue
\mciteSetBstMidEndSepPunct{\mcitedefaultmidpunct}
{\mcitedefaultendpunct}{\mcitedefaultseppunct}\relax
\EndOfBibitem
\bibitem[Watson and Chan(2012)Watson, and Chan]{watson_excited_2012}
Watson,~M.~A.; Chan,~G. K.-L. Excited {States} of {Butadiene} to {Chemical}
  {Accuracy}: {Reconciling} {Theory} and {Experiment}. \emph{J. Chem. Theory
  Comput.} \textbf{2012}, \emph{8}, 4013--4018\relax
\mciteBstWouldAddEndPuncttrue
\mciteSetBstMidEndSepPunct{\mcitedefaultmidpunct}
{\mcitedefaultendpunct}{\mcitedefaultseppunct}\relax
\EndOfBibitem
\bibitem[Daday \latin{et~al.}(2012)Daday, Smart, Booth, Alavi, and
  Filippi]{daday_full_2012}
Daday,~C.; Smart,~S.; Booth,~G.~H.; Alavi,~A.; Filippi,~C. Full {Configuration}
  {Interaction} {Excitations} of {Ethene} and {Butadiene}: {Resolution} of an
  {Ancient} {Question}. \emph{J. Chem. Theory Comput.} \textbf{2012}, \emph{8},
  4441--4451\relax
\mciteBstWouldAddEndPuncttrue
\mciteSetBstMidEndSepPunct{\mcitedefaultmidpunct}
{\mcitedefaultendpunct}{\mcitedefaultseppunct}\relax
\EndOfBibitem
\bibitem[Widmark \latin{et~al.}(1990)Widmark, Malmqvist, and
  Roos]{widmark_density_1990}
Widmark,~P.-O.; Malmqvist,~P.-{\AA}.; Roos,~B.~O. Density matrix averaged
  atomic natural orbital ({ANO}) basis sets for correlated molecular wave
  functions. \emph{Theor. Chim. Acta} \textbf{1990}, \emph{77}, 291--306\relax
\mciteBstWouldAddEndPuncttrue
\mciteSetBstMidEndSepPunct{\mcitedefaultmidpunct}
{\mcitedefaultendpunct}{\mcitedefaultseppunct}\relax
\EndOfBibitem
\end{mcitethebibliography}
\end{document}